\newif\ifmnras

\ifmnras

\documentclass[a4paper,fleqn,usenatbib]{mnras}
\usepackage{newtxtext,newtxmath}
\usepackage[T1]{fontenc}
\usepackage{ae,aecompl}

\usepackage{graphicx}	
\usepackage[section] {placeins}
\usepackage{subfigure}
\usepackage{float}
\usepackage{color}
\usepackage{hyperref}
\graphicspath{{./figures/}}
\usepackage{float}

\def\msun{{\rm\,M_\odot}} 
\def\lsun{{\rm\,L_\odot}}
\def\zsun{{\rm\,Z_\odot}}
\newcommand{\etal}{et al.\ }
\newcommand{\kms}{\, {\rm km\, s}^{-1}}
\newcommand{\ikms}{(\kms)^{-1}}
\newcommand{\mpc}{\, {\rm Mpc}}
\newcommand{\kpc}{\, {\rm kpc}}
\newcommand{\hmpc}{\, h^{-1} \mpc}
\newcommand{\ihmpc}{(\hmpc)^{-1}}
\newcommand{\hkpc}{\, h^{-1} \kpc}
\newcommand{\lya}{Ly$\alpha$}
\newcommand{\lyaf}{Ly$\alpha$ forest}
\newcommand{\ch}{\bf change}
\newcommand{\gmo}{{\gamma-1}}
\newcommand{\bF}{\bar{F}}
\newcommand{\hi}{\mbox{H\,{\scriptsize I}\ }}
\newcommand{\heii}{\mbox{He\,{\scriptsize II}\ }}
\newcommand{\civ}{\mbox{C\,{\scriptsize IV}\ }}
\newcommand{\kpa}{k_\parallel}
\newcommand{\vk}{{\mathbf k}}
\newcommand{\df}{\delta_F}
\newcommand{\sF}{{F_s}}
\newcommand{\sdelta}{{\delta_s}}
\newcommand{\seta}{{\eta_s}}
\newcommand{\dt}{\Delta \theta}
\newcommand{\dv}{\Delta v}
\newcommand{\pa}{\parallel}
\newcommand{\pe}{\perp}
\newcommand{\dz}{\Delta z}
\newcommand{\llya}{L$_{{\rm Ly}\alpha}$}
\newcommand{\lheii}{L$_{{\rm He II}}$}
\newcommand{\lciv}{L$_{{\rm C IV}}$}
\newcommand{\expZ}{$\langle Z \rangle$}
\newcommand{\expT}{$\langle T \rangle$}
\newcommand{\expD}{$\langle n_{{\rm H}} \rangle$}
\newcommand{\expF}{$\langle f_{{\rm HI}} \rangle$}
\def\h2{${\rm\,H_2}$}

\author[]{Renyue Cen
\\
Department of Astrophysical Sciences, Princeton University, Princeton, NJ 08544\\
} 

\pubyear{2015}

\begin{document}
\label{firstpage}
\pagerange{\pageref{firstpage}--\pageref{lastpage}}
\maketitle

\else

\pdfoutput=1
\documentclass[12pt,preprint]{aastex}

\textheight=9.2in
\topmargin=-0.5in
\textwidth=6.5in
\rightmargin=2.0in

\usepackage[T1]{fontenc}
\usepackage{ae,aecompl}

\usepackage{graphicx}	
\usepackage[section] {placeins}
\usepackage{subfigure}
\usepackage{float}
\usepackage{color}
\usepackage{hyperref}
\graphicspath{{./figures/}}
\usepackage{float}

\usepackage[scaled]{helvet}
\renewcommand*\familydefault{\sfdefault}
\usepackage[T1]{fontenc}

\usepackage{amsmath}  

\def\msun{{\rm\,M_\odot}} 
\def\lsun{{\rm\,L_\odot}}
\def\zsun{{\rm\,Z_\odot}}
\newcommand{\etal}{et al.\ }
\newcommand{\kms}{\, {\rm km\, s}^{-1}}
\newcommand{\ikms}{(\kms)^{-1}}
\newcommand{\mpc}{\, {\rm Mpc}}
\newcommand{\kpc}{\, {\rm kpc}}
\newcommand{\hmpc}{\, h^{-1} \mpc}
\newcommand{\ihmpc}{(\hmpc)^{-1}}
\newcommand{\hkpc}{\, h^{-1} \kpc}
\newcommand{\lya}{Ly$\alpha$}
\newcommand{\lyaf}{Ly$\alpha$ forest}
\newcommand{\ch}{\bf change}
\newcommand{\gmo}{{\gamma-1}}
\newcommand{\bF}{\bar{F}}
\newcommand{\hi}{\mbox{H\,{\scriptsize I}\ }}
\newcommand{\heii}{\mbox{He\,{\scriptsize II}\ }}
\newcommand{\civ}{\mbox{C\,{\scriptsize IV}\ }}
\newcommand{\kpa}{k_\parallel}
\newcommand{\vk}{{\mathbf k}}
\newcommand{\df}{\delta_F}
\newcommand{\sF}{{F_s}}
\newcommand{\sdelta}{{\delta_s}}
\newcommand{\seta}{{\eta_s}}
\newcommand{\dt}{\Delta \theta}
\newcommand{\dv}{\Delta v}
\newcommand{\pa}{\parallel}
\newcommand{\pe}{\perp}
\newcommand{\dz}{\Delta z}
\newcommand{\llya}{L$_{{\rm Ly}\alpha}$}
\newcommand{\lheii}{L$_{{\rm He II}}$}
\newcommand{\lciv}{L$_{{\rm C IV}}$}
\newcommand{\expZ}{$\langle Z \rangle$}
\newcommand{\expT}{$\langle T \rangle$}
\newcommand{\expD}{$\langle n_{{\rm H}} \rangle$}
\newcommand{\expF}{$\langle f_{{\rm HI}} \rangle$}
\def\h2{${\rm\,H_2}$}

\title{Constraint on Matter Power Spectrum on $10^6-10^9\msun$ Scales from ${\large\tau_e}$}
\author{Renyue Cen$^{1}$} 

\begin{document}
\label{firstpage}

\fi

\begin{abstract}

An analysis of the physics-rich endgame of
reionization at $z=5.7$ is performed, utilizing jointly the observations of the Ly$\alpha$ forest,
the mean free path of ionizing photons, the luminosity function of galaxies and new physical insight.
We find that an upper limit on ${\rm \tau_e}$ provides a constraint on 
the minimum mean free path (of ionizing photons) that is primarily due to dwarf galaxies, 
which in turn yields a new and yet the strongest 
constraint on the matter power spectrum on $10^6-10^9\msun$ scales.
With the latest Planck measurements of ${\rm \tau_e = 0.055 \pm 0.009}$,
we can place 
an upper limit of $(8.9\times 10^6, 3.8\times 10^7, 4.2\times 10^8)\msun$ 
on the lower cutoff mass of the halo mass function,
or equivalent a lower limit on warm dark matter particle mass ${\rm m_x \ge (15.1, 9.8, 4.6)keV}$
or on sterile neutrino mass ${\rm m_s \ge (161, 90, 33)keV}$, at $(1, 1.4, 2.2)\sigma$
confidence level, respectively.

\end{abstract}






\section{Introduction}

The \citet[][]{1965Gunn} optical depth of Ly$\alpha$ photons 
provides the strongest and most sensitive constraint on the neutral hydrogen fraction 
of the intergalactic medium (IGM).
The integrated electron scattering optical depth of the universe 
provides a complementary constraint on the ionized fraction of the IGM, but 
is insensitive to the neutral hydrogen fraction as long as the IGM is mostly ionized.

Recent measurements of the electron scattering optical depths
of the IGM by the cosmic microwave background radiation experiments
\citep[e.g.,][]{2013Hinshaw, 2015Planck} suggest
that it may be significantly below redshift $z=12$ before the universe becomes half reionized.
The observations of the 
high redshift ($z>6$) quasar absorption spectra from the Sloan Digital Sky Survey (SDSS) and others
\citep[e.g.,][]{2006Fan} and arguments based on the slowly and continuously evolving IGM opacity
\citep[e.g.,][]{2007Becker} suggest that only 
at $z=5.7$ the universe is sufficiently ionized to allow for detectable transmission of Ly$\alpha$ photons
hence definitive measurements of (low enough) Ly$\alpha$ (and higher order Lyman series) optical depth.

It is generally accepted that stars are
primarily responsible for producing most of the ionizing photons for cosmological reionization.
While it seems relatively secure to further suggest that the
reionization process has begun at $z\ge 10$ based on analysis of expected emergence of first galaxies
in the standard cold dark matter model \citep[e.g.,][]{2015Trac},
the combination of these independent observational indications
now paints a reionization picture that is rapidly evolving at $z=6-10$.
Two important implications are that the so-called first galaxies that form out of primordial gas 
may be closer to us than thought before and that Popolation III (Pop III) stars 
formed with metal-free gas may extend to more accessible redshifts.

In this contribution we perform a detailed analysis of the endgame of
the cosmological reionization at $z=5.7$.
We examine joint constraints on the IGM from considerations of both global and local ionization balances
observationally and, for the first time, 
self-consistently in the context of the standard cold dark matter model.
We find reasonable concordance between Ly$\alpha$ optical depth, Lyman continuum (LyC) mean free path (mfp)
$\lambda_{\rm mfp}$ and global recombination rate of hydrogen
observationally and theoretically.
We solve the global reionization equation, 
given the emissivity evolution in the context of the standard cold dark matter 
model normalized to the boundary 
conditions of required emissivity at $z=5.7$ and reionization completing at $z=5.7$.
We provide a detailed analysis of the attainable solutions of reionization histories
to shed light on the overall topological evolution of the HII regions,
the evolution of the Ly$\alpha$ emitters,
the neutral fraction of the IGM,
and a new and powerful constraint on the matter power spectrum on small scales
hence dark matter particle properties.

Our focus here is on placing a yet the strongest constraint on the scale-scale power in the cosmological model and,
specifically, the strongest lower bound on the mass of warm dak matter particles.
The physical insight on this particular point is new and may be described briefly as follows.
The state of the IGM at $z=5.7$ is well fixed by 
the \citet[][]{1965Gunn} optical depth of Ly$\alpha$ photons, which in turn provides
a tight constraint on the photoionization rate $\Gamma$ at $z=5.7$ in the post-reionization epoch.
Since $\Gamma$ at $z=5.7$ is equal to  ${\rm \dot N_{ion,IGM} \lambda_{\rm mfp}\bar\sigma_{ion}}$,
where 
${\rm \dot N_{ion,IGM}}$ is the global mean of effective ionization photon emissivity at $z=5.7$,
${\rm \lambda_{\rm mfp}}$ is the mean free path of ionizing photons at $z=5.7$
and ${\rm \bar\sigma_{ion}}$ is the spectrum-weighted mean photoionization cross section, a constant.
Thus, a tight constraint on  
$\Gamma$ at $z=5.7$ is equivalent to an equally tight constraint on the product 
${\rm \dot N_{ion,IGM} \lambda_{\rm mfp}}$ at $z=5.7$.
Note that ${\rm \dot N_{ion,IGM}}$ already takes into account the escape fraction of ionizing photon from ionization sources
(e.g., galaxies and others).
The degeneracy between ${\rm \dot N_{ion,IGM}}$ and ${\rm \lambda_{\rm mfp}}$ 
can be broken, if one considers, jointly, a separate constraint placed by 
an upper limit on the integrated electron scattering optical depth of the universe 
${\rm \tau_e}$ from the latest cosmic microwave background radiation experiments \citep[e.g.,][]{2016Planck}.
This is where our new physical insight comes in.
We point out that, when the product ${\rm \dot N_{ion,IGM} \lambda_{\rm mfp}}$ is fixed,
a higher $\lambda_{\rm mfp}$ would require a lower ${\rm \dot N_{ion,IGM}}$,
which in turn would cause the reionization process to shift to lower redshift hence give rise to a lower $\tau_e$.
In other words, there is a negative correlation between $\lambda_{\rm mfp}$ and $\tau_e$.
Since more small-scale power results in a lower $\lambda_{\rm mfp}$,
there is then a negative correlation between the amount of small-scale power and $\tau_e$
- more small-scale power leads to lower $\tau_e$.
As a result, an upper bound on $\tau_e$ placed by the latest CMB observations
would translate to a lower bound on the amount of small-scale power hence a lower bound
on the particle mass in the context of the warm dark matter model.
This is the scientific focus of this paper.

\section{On Sinks and Sources of Lyman Continuum at $z=5.7$}

\subsection{Global Balance of Emission and Recombination}

The hydrogen recombination rate
per unit comoving volume at redshift $z$ is
\begin{equation}
{\rm \dot N_{rec} = C_{H II} \alpha_B(T) [1+Y_p/4(1-Y_p)] n_{H,0}^2 (1+z)^3}
\label{eq:Nion}
\end{equation}
and the corresponding helium I recombination rate is
\begin{equation}
{\rm \dot N_{HeI, rec} = C_{H II} \alpha_B(HeI,T) [1+Y_p/4(1-Y_p)] [Y_p/4(1-Y_p)] n_{H,0}^2 (1+z)^3},
\label{eq:NionHeI}
\end{equation}
where ${\rm n_{H,0} = 2.0\times 10^{-7} (\Omega_B/0.048) cm^{-3}}$ is
the mean hydrogen number density at $z=0$,
${\rm Y_p=0.24}$ the primordial helium mass fraction,
${\rm C_{H II}}$ is the clumping factor of the recombining medium.
The case B recombination coefficient \\
${\rm \alpha_B(T)  = (2.59,2.52)\times 10^{-13}}~$cm$^3$s$^{-1}$ at ${\rm T=(10^4,2\times 10^4)}$K
\citep[][]{1989Osterbrock}.
The case B He I recombination coefficient is 
${\rm \alpha_B (HeI, T)  = (2.73,1.55)\times 10^{-13}~cm^3~s^{-1}}$ at ${\rm T= (10^4,2\times 10^4)}$K
\citep[][]{1989Osterbrock}.

To prevent the already ionized IGM from recombining,
the amount of ionizing photons entering the IGM has to be, at least, equal to 
the total recombination rate, resulting in the well known minimum 
requirement of ionizing photon production rate \citep[e.g.,][]{1999Madau} 
\begin{equation} \label{eq:Nionglobal}
\begin{split}
{\rm \dot N_{ion,global}} &{\rm \ge \dot N_{rec} + \dot N_{HeI,rec}}  \\ 
&{\rm = 3.4\times 10^{50}(C_{H II}/3.2)(\Omega_b/0.048)^2 ((1+z)/6.7)^3 cMpc^{-3} s^{-1}\ for\ T=10^4K} \\
&{\rm = 3.2\times 10^{50}(C_{H II}/3.2)(\Omega_b/0.048)^2 ((1+z)/6.7)^3 cMpc^{-3} s^{-1}\ for\ T=2\times 10^4K},
\end{split}
\end{equation}
\noindent
assuming that helium II is not ionized.
We shall call this constraint expressed in 
Eq \ref{eq:Nionglobal} ``global constraint".
For clarity we will adopt the convention to use 
${\rm cMpc}$ and ${\rm pMpc}$ to denote comoving 
and proper ${\rm Mpc}$, respectively.
Early hydrodynamical simulations suggest 
${\rm C_{HII} \sim 10-40}$ at $z<8$ \citep[e.g.,][]{1997Gnedin}.
More recent simulations that separate out dense
interstellar medium (ISM) from the IGM
indicate a lower ${\rm C_{HII} \sim 1-6}$ at $z\sim 6$
\citep[e.g.,][]{2003Sokasian, 2006Iliev, 2009Pawlik, 2012Shull, 2012Finlator}.
\citet[][]{2009Pawlik} give
\begin{equation} \label{eq:CHII}
\begin{split}
{\rm C_{HII}} &={\rm 3.2\quad for \quad z\le 10} \\ 
&={\rm 1+\exp{(-0.28z+3.59)}\quad for \quad z>10},
\end{split}
\end{equation}
\noindent
which we will use in the calculations below.
As we demonstrate later, the value ${\rm C_{HII}=3.2}$ at $z=5.7$
is concordant between considerations of global and local ionization balances.

\subsection{Local Balance of Ionization and Recombination}

A second, independent determination of ionizing photon production rate can be 
obtained from the Ly$\alpha$ optical depth around cosmic mean density, ${\rm \tau_{Ly\alpha}}$,
i.e., the \citet[][]{1965Gunn} optical depth, at $z=5.7$, where observational measurements are available.
Because of the large cross section of neutral hydrogen for Ly$\alpha$ scattering, 
${\rm \tau_{Ly\alpha}}$
is the most sensitive probe of neutral medium
in the low neutral-fraction regime.
From the SDSS observations of high redshift quasar absorption spectra 
${\rm \tau_{Ly\alpha}}$ is directly measured \citep[][]{2002Fan, 2006Fan}.
When analyzed in conjunction with density distributions 
of the IGM from hydrodynamic simulations, 
one can infer both the volume weighted neutral fraction and 
the ionization rate $\Gamma$, expressed in units of ${\rm 10^{-12} s^{-1}}$, $\Gamma_{-12}$. 
Because the mean density regions that determine the volume-weighted neutral fraction
are well resolved in simulations (i.e., the simulation resolution is much finer
than the Jeans scale of the photoionized IGM),
the uncertainty on the determined volume-weighted neutral fraction is small
and does not depend sensitively on cosmological parameters, either.
The analysis performed by \citet[][]{2002Cen} 
uses a smaller sample of SDSS quasars coupled with simulations of \citet[][]{1994Cen}.
The analysis performed by \citet[][]{2006Fan} 
utilizes a larger quasars sample and  
the density distribution function of \citet[][]{2000MiraldaEscude}. 
Both studies derive, independently, $\Gamma_{-12}\sim 0.20$.
For the subsequent calculations, we will use 
\begin{equation}
{\rm \Gamma_{-12}=0.20_{-0.06}^{+0.11}}
\label{eq:gamma12}
\end{equation}
\noindent
at $z=5.7$ from \citet[][]{2006Fan}.

Under the assumption that the spatial scales of fluctuations (or clustering scales)
for both sources and sinks are substantially smaller than the mean free path $\lambda_{\rm mfp}$ of LyC photons,
then the (approximately uniform) ionizing flux at any spatial point is
\begin{equation}
{\rm F_{ion} = \int_0^\infty {\dot N_{ion,IGM}\over 4\pi r^2} \exp{(-r/\lambda_{\rm mfp})}4\pi r^2 dr = \dot N_{ion,IGM} \lambda_{\rm mfp}},
\label{eq:flux}
\end{equation}
\noindent
where ${\rm \dot N_{ion,IGM}}$ is the mean emissivity of ionizing photons entering
the IGM.
We note that the 2-point correlation length 
of galaxies at $z=5.7$ is $4-5$cMpc \citep[e.g.,][]{2010Ouchi},
much smaller than $\lambda_{\rm mfp}\sim 30-60$cMpc, which we will discuss later.
Therefore, the above assumption is a good one, so long as stellar sources 
are the main driver of cosmological reionization.
We expect that radiation flux fluctuations would be on the order of the ratio
of the two lengths scales above, i.e., $\sim 10\%$.
As we will show later that, in the context of the $\Lambda$CDM model,
$\lambda_{\rm mfp}$ depends on $\Gamma$ approximately as $\lambda_{\rm mfp}\propto \Gamma^{-0.28}$.
Thus, we expect that the uniform radiation assumption is accurate statistically for computing 
the mean $\lambda_{\rm mfp}$ at $1-3\%$ level, with negligible systematic biases.
The hydrogen ionization rate 
\begin{equation}
{\rm \Gamma = F_{ion}\bar\sigma_{ion} = \dot N_{ion,IGM} \lambda_{\rm mfp}\bar\sigma_{ion}},
\label{eq:gamma}
\end{equation}
\noindent
where $\bar\sigma_{ion}$ is the spectrum-weighted mean photoionization cross section, 
\begin{equation}
{\rm \bar\sigma_{ion} \equiv 
{{\int_{13.6eV}^\infty{f_\nu\over h\nu} \sigma_H(\nu)d\nu}
\over
{\int_{13.6eV}^\infty{f_\nu\over h\nu}d\nu}
}},
\label{eq:sigmaH}
\end{equation}
\noindent
where ${\rm \sigma_H(\nu)}$ is the photon energy-dependent hydrogen ionization cross section,
${\rm f_\nu}$ is the ionizing photon spectrum.
We will use ${\rm f_\nu}$ for Pop II stars of metallicity $Z=0.05\zsun$ from \citet[][]{2001Tumlinson},
which may be approximated as  
\begin{alignat*}{2} \label{eq:fnu}
{\rm f_\nu} &{\rm \propto \nu^{0}} &&{\rm\quad for\quad \nu=13.6-24.6eV}\\ 
&{\rm \propto \nu^{-1}} &&{\rm \quad for\quad \nu=24.6-46eV} \\
&{\rm \propto \nu^{-\infty}} &&{\rm \quad for\quad \nu>46eV},
\end{alignat*}
\noindent
which results in the fiducial value that we will use in our calculations at $z=5.7$, 
\begin{equation}
{\rm \bar\sigma_{ion} = 3.16\times 10^{-18}~cm^{2}}.
\label{eq:sigmaion}
\end{equation}
\noindent
Combining Eq (\ref{eq:gamma12},
\ref{eq:gamma},
\ref{eq:sigmaion}) 
gives the constraint on comoving emissivity at $z=5.7$ from Gunn-Peterson optical depth, named "local constraint",
\begin{equation} 
{\rm \dot N_{ion,local} = 2.7\times 10^{50} ({\Gamma_{-12}\over 0.2}) ({\bar\sigma_{ion}\over 3.16\times 10^{-18} cm^{2}})^{-1}({\lambda_{mfp}\over 7.6pMpc})^{-1} cMpc^{-3} s^{-1}}.
\label{eq:Nionlocal}
\end{equation}
\noindent
In Eq \ref{eq:Nionlocal} it is seen that
there is a significant, linearly inverse dependence of 
${\rm \dot N_{ion,local}}$ on ${\rm \lambda_{mfp}}$, which we now discuss in length observationally 
here and theoretically in the next subsection.

Traditionally, ${\rm \lambda_{mfp}}$ 
is determined by counting the incidence frequency of Lyman limit systems (LLSs)
\citep[e.g.,][]{1994StorrieLombardi, 
1995StenglerLarrea,
2010Songaila,
2011bRibaudo,
2013OMeara} and generally found to be in the range of ${\rm \lambda_{mfp}=5-10~pMpc}$ at $z=5.7$,
when extrapolated from lower redshift trends.
This method to determine ${\rm \lambda_{mfp}}$
contains some ambiguity as to the dependence 
of the incidence frequency on exact choice of column density threshold of LLSs,
and uncertainties 
related to absorption system identifications (such as line blending)
and collective absorption due to clustering of absorbers.
A more direct approach to determining ${\rm \lambda_{mfp}}$ 
is to measure the optical depth at Lyman limit directly,
as pioneered by \citet[][]{2009bProchaska}.
A recent application of that technique to 
a large sample of (163) high redshift quasars is cast into fitting formula
${\rm \lambda_{mfp}=37[(1+z)/5]^{-5.4\pm 0.4}}$pMpc that covers up to redshift $z=5.5$
\citep[][]{2014bWorseck}.
Extrapolating this formula to $z=5.7$ results in a median value of $7.6$~pMc, 
\begin{equation} 
{\rm \lambda_{mfp} = 7.6_{-0.8}^{+1.0}~pMpc},
\label{eq:lambdamfpLLS}
\end{equation}
\noindent
with the $1$ and $2\sigma$ range of $6.8-8.6~$pMpc and $6.0-9.6~$pMpc, respectively.
It is seen that the directly measured 
${\rm \lambda_{mfp}}$ are in broad agreement with those based on counting LLSs,
which is reassuring. 
Nevertheless, it is prudent to bear in mind a significant caveat that 
${\rm \lambda_{mfp}}$ at $z=5.7$ is not directly observed but requires extrapolation from lower redshift data.

\subsection{Concordance of Independent Observations at $z=5.7$}

\ifmnras

\begin{figure}
\centering
\includegraphics[width=3.5in, keepaspectratio]{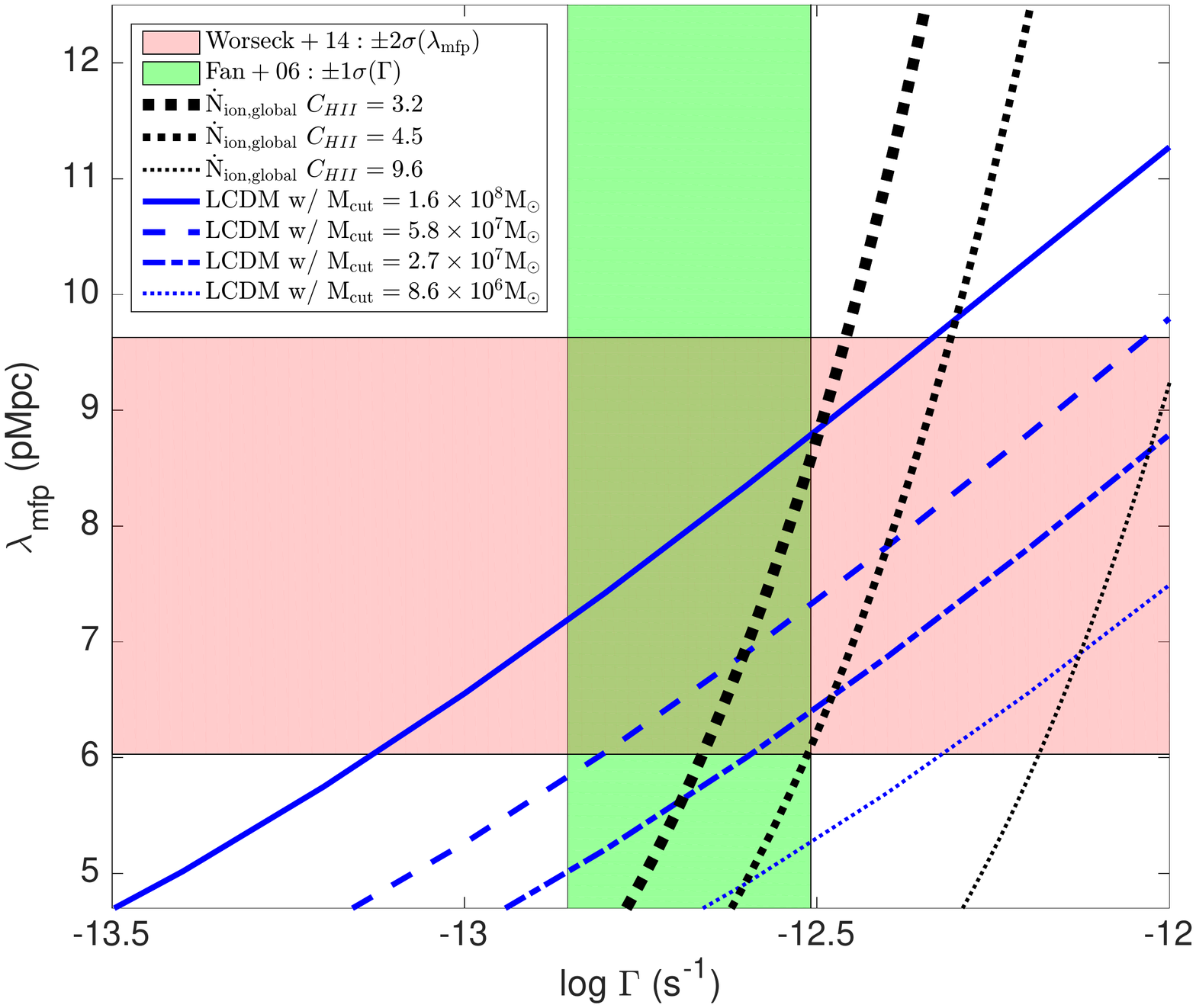}
\caption{
{\color{red}Left panel:}
shows four independent sets of constraints on the ${\rm \Gamma-\lambda_{mfp}}$ plane:
(1) the observed ${\rm \pm 2\sigma}$ range for ${\rm \lambda_{mfp}}$ from \citet[][]{2014bWorseck} based on LyC 
optical depth observed at $z<5.5$ extrapolate to $z=5.7$ (see Eq \ref{eq:lambdamfpLLS})
shown as the red shaded horizontal region;
(2) the observationally inferred $1\sigma$ range of $\Gamma$ based on 
measurement of Ly$\alpha$ absorption optical depth at $z=5.7$
from \citet[][]{2006Fan} shown as the green shaded vertical region (see Eq \ref{eq:gamma12});
(3) lower bound based on a global balance between emissivity and recombination 
with Eq \ref{eq:Nionglobal} assuming clumping factor ${\rm C_{HII}=(3.2,4.5,9.6)}$ and 
gas temperature $T=10^4~$K, shown as dotted black (thick, median thick, thin) curves;
(4) the self-consistently calculated relation between 
${\rm \Gamma}$ and ${\rm \lambda_{mfp}}$ in the standard $\Lambda$CDM model
with a lower halo mass cutoff 
of $(1.6\times 10^8, 5.8\times 10^7, 2.7\times 10^7, 8.6\times 10^6)\msun$,
shown as (solid, dashed, dot-dashed, dotted) curves, respectively.
}
\label{fig:bb}
\end{figure}

\else

\begin{figure}[!h]
\begin{center}
\includegraphics[width=5.9in, keepaspectratio]{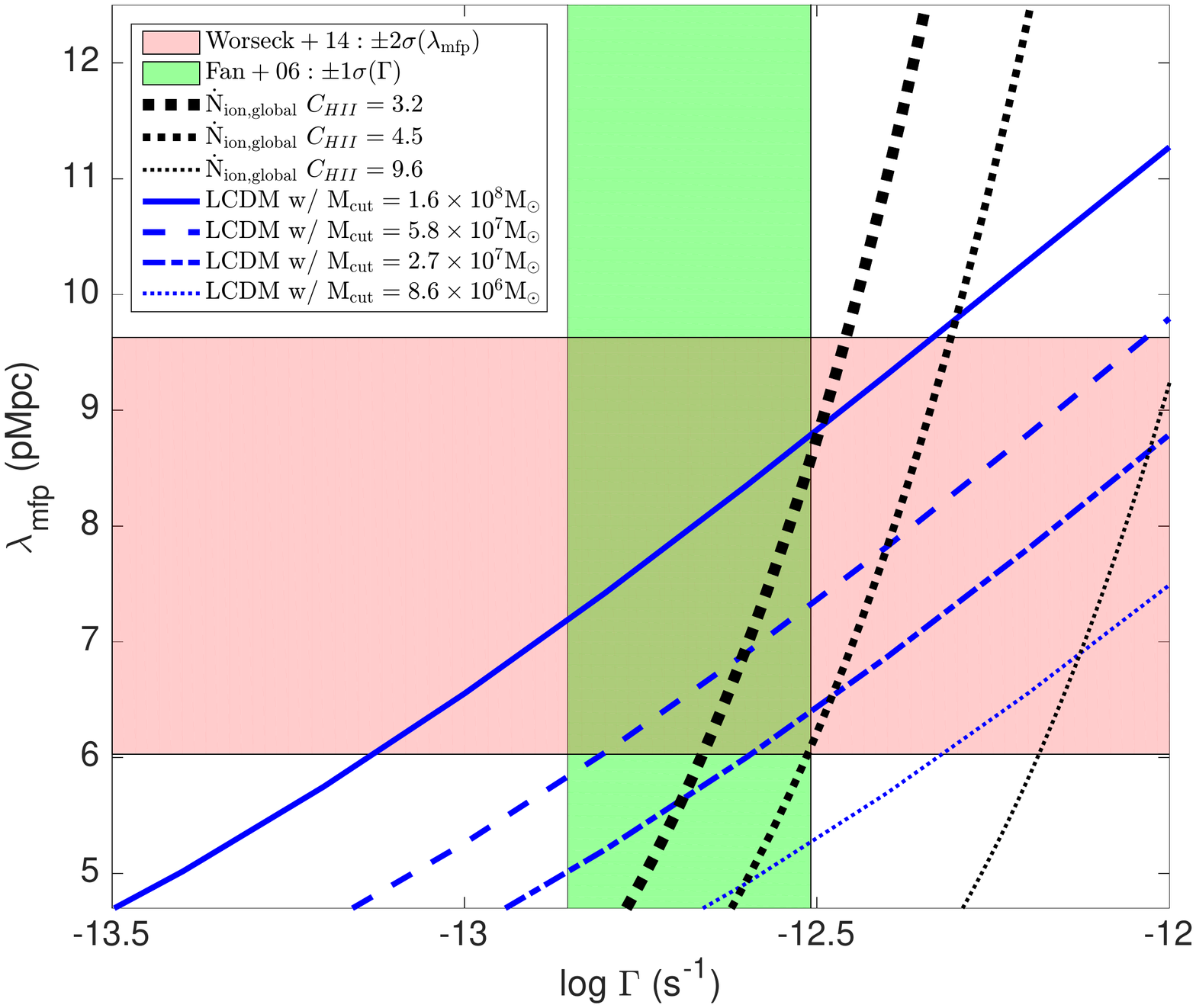}
\end{center}
\vskip -1.0cm
\caption{
shows four independent sets of constraints on the ${\rm \Gamma-\lambda_{mfp}}$ plane:
(1) the observed ${\rm \lambda_{mfp}}$ from \citet[][]{2014bWorseck} based on LyC 
optical depth observed at $z<5.5$ extrapolate to $z=5.7$ (see Eq \ref{eq:lambdamfpLLS})
shown as the red solid curve (mean),
thick red dashed curves ($1\sigma$) and thin red dashed curves ($2\sigma$);
(2) the observationally inferred $1\sigma$ range of $\Gamma$ based on 
measurement of Ly$\alpha$ absorption optical depth at $z=5.7$
from \citet[][]{2006Fan} shown as the two vertical green dashed lines
(see Eq \ref{eq:gamma12});
(3) lower bound based on a global balance between emissivity and recombination 
with Eq \ref{eq:Nionglobal} assuming clumping factor ${\rm C_{HII}=(3.2,4.5,9.6)}$ and 
gas temperature $T=10^4~$K, shown as dotted black (thick, median thick, thin) curves;
(4) the self-consistently calculated relation between 
${\rm \Gamma}$ and ${\rm \lambda_{mfp}}$ in the standard $\Lambda$CDM model
with a lower halo mass cutoff 
of $(1.6\times 10^8, 5.8\times 10^7, 2.7\times 10^7, 8.6\times 10^6)\msun$,
respectively, corresponding to a virial temperature cutoff of ${\rm T_{v,cutoff}=(10^4,5\times 10^3, 3\times 10^3, 1.4\times 10^3})$K.
}
\label{fig:mfp}
\end{figure}

\fi

We now combine three independent sets of observational 
constraints on ${\rm \dot N_{ion}}$, $\Gamma$ and ${\rm \lambda_{mfp}}$ 
on the ${\rm \Gamma-\lambda_{mfp}}$ plane, shown in Figure \ref{fig:mfp}: 
(1) the observed ${\rm \lambda_{mfp}}$ from \citet[][]{2014bWorseck} based on Lyman continuum 
radiation optical depth at $z=5.7$ (see Eq \ref{eq:lambdamfpLLS})
are shown as the red solid curve (mean),
thick red dashed curves ($1\sigma$) and thin red dashed curves ($2\sigma$);
(2) the observationally inferred $1\sigma$ range of $\Gamma$ based on 
measurement of Ly$\alpha$ absorption optical depth at $z=5.7$
from \citet[][]{2006Fan} are shown as the two vertical green dashed lines
(see Eq \ref{eq:gamma12});
(3) lower bound based on a global balance between emissivity and recombination 
with Eq \ref{eq:Nionglobal} assuming clumping factor ${\rm C_{HII}=(3.2,4.5,9.6)}$ and 
gas temperature $T=10^4~$K, shown as dotted black (thick, median thick, thin) curves.

To be conservative, we will use the $2\sigma$ range of 
${\rm \lambda_{mfp}}$ from \citet[][]{2014bWorseck} for our discussion, 
because of the possible additional, systematic uncertainty of using an extrapolated 
value from the observed highest redshift of $z=5.5$ to $z=5.7$.
Thus, the allowed parameter space is enclosed by the two thin dashed red horizontal lines
and the two vertical dashed green lines.
This space is then further constrained by the requirement that only to the right of 
each of the dotted black curves is attainable, depending on the assumed clumpying factor ${\rm C_{HII}}$.
The placement of this additional requirement on the plane 
suggests that ${\rm C_{HII}>5}$ at $z=5.7$  may not be feasible but the values 
in Eq \ref{eq:CHII} that is obtained from recent radiation hydrodynamic simulations 
and adopted here are fully consistent with this constraint.

It is by no means guaranteed a priori that there is any parameter space left 
when all these three independent observational constraints are considered,
due to uncertainties in individual observations.
Hence, the fact that there is suggests a concordance among the independent observations.

\subsection{Global Stellar Emissivity of Ionizing Photons at $z=5.7$}

Figure \ref{fig:mfp} in \S 2.4 summarizes the current state of constraints on the required 
emissivity of ionizing photons in the IGM at $z=5.7$, in order to (1) keep the IGM ionized globally,
(2) keep the IGM ionized locally as demanded by the optical depths probed by the hydrogen Lyman series 
absorption lines.
The multi-faceted agreement is indeed quite remarkable, providing 
a validation of the different observations at $z=5.7$ (in some cases extrapolation is needed) 
in the post-overlap epoch.

\ifmnras

\begin{figure}
\centering
\includegraphics[width=3.5in, keepaspectratio]{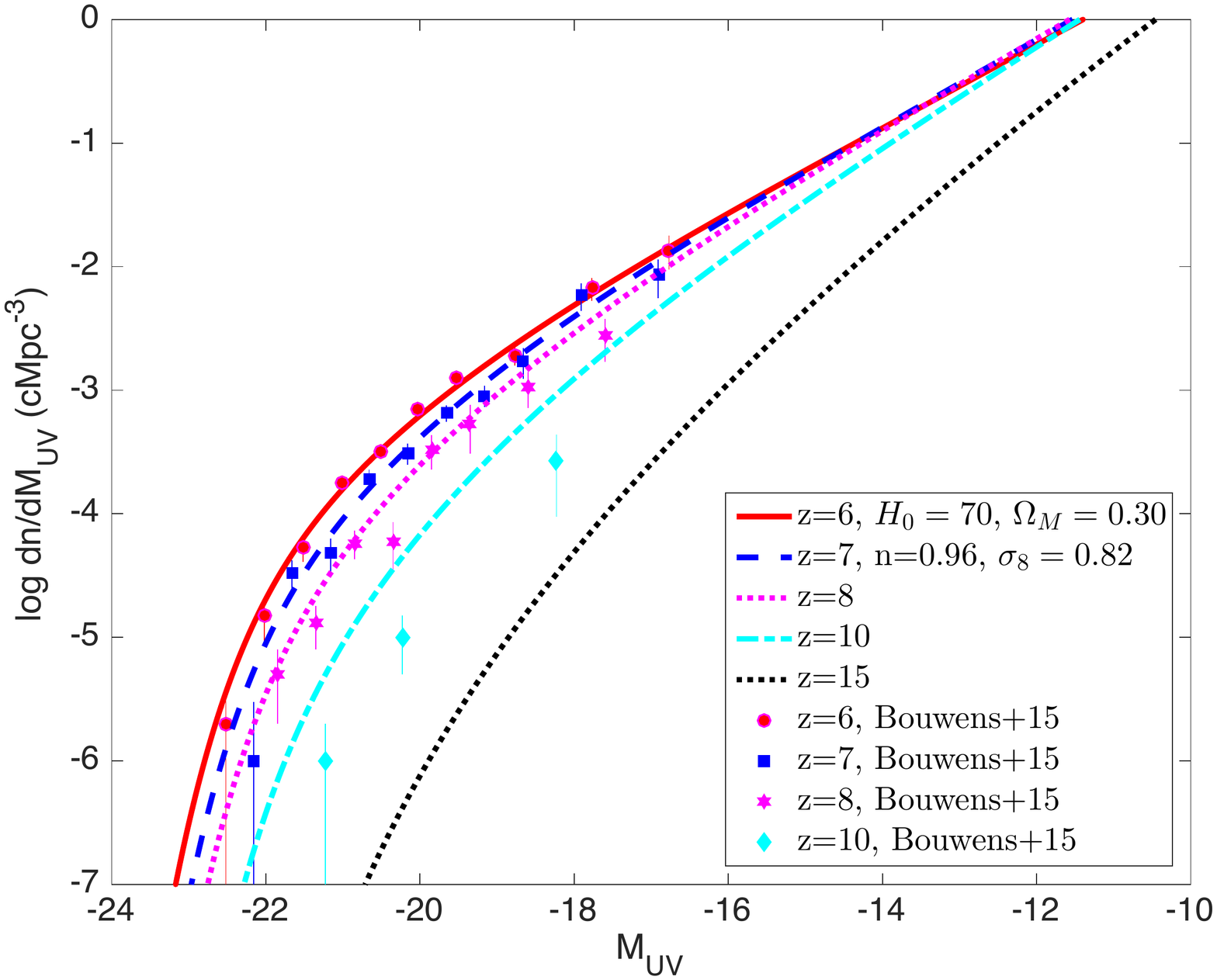}
\caption{
{\color{red}Left panel:}
shows the galaxy luminosity functions predicted by the $\Lambda$CDM model
at $z=6$ (red solid curve), 7 (blue dashed curve), 8 (magenta dotted curve), 10 (cyan dot-dashed curve)
and 15 (black dotted curve),
which are compared to the observations at the four corresponding redshifts,
shown as various symbols with corresponding colors.
The observational data are from \citet[][]{2015Bouwens}.
}
\label{fig:bb}
\end{figure}

\else

\begin{figure}[!h]
\begin{center}
\includegraphics[width=5.5in, keepaspectratio]{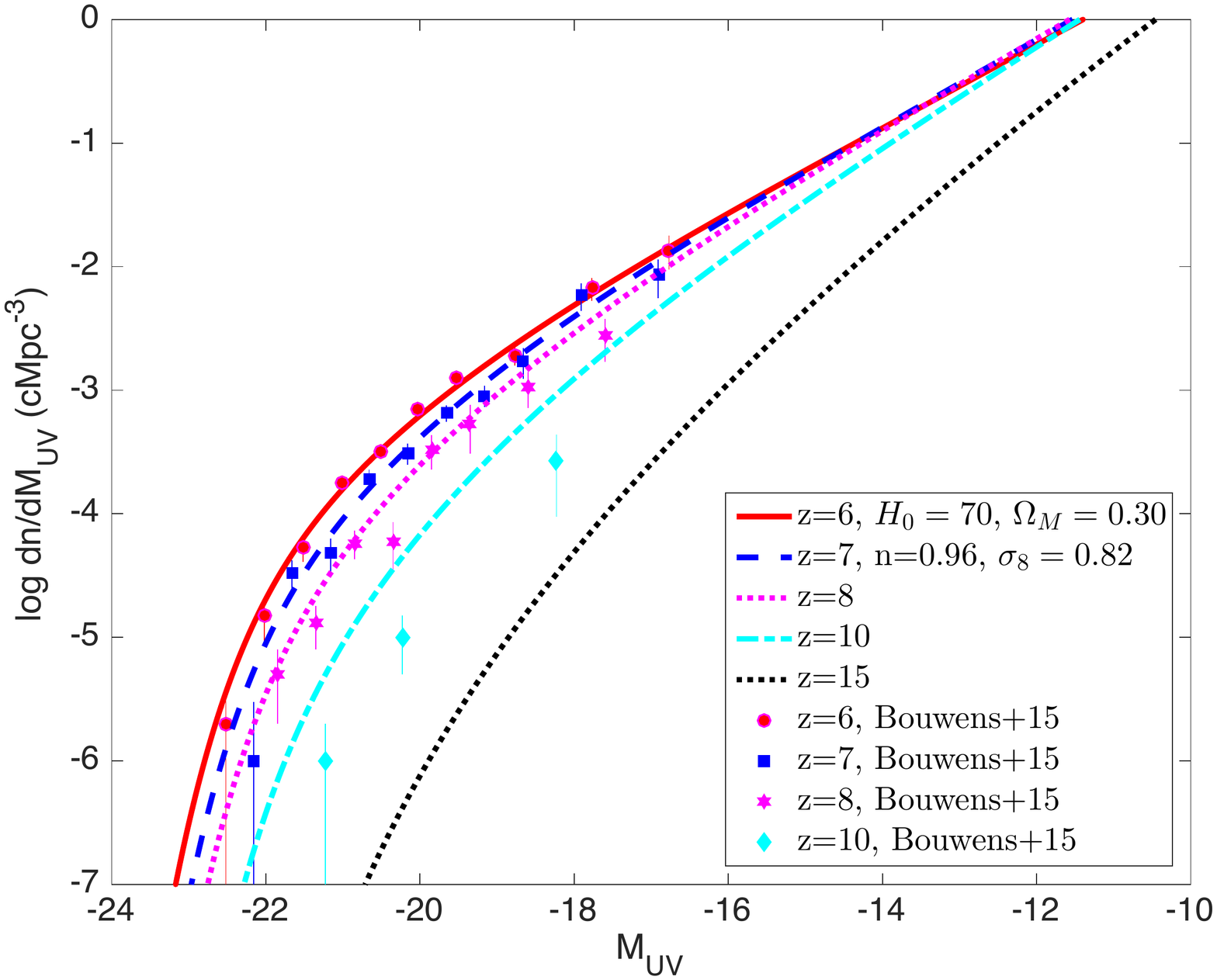}
\end{center}
\vskip -1.0cm
\caption{
shows the galaxy luminosity functions predicted by the $\Lambda$CDM model
at $z=6$ (red solid curve), 7 (blue dashed curve), 8 (magenta dotted curve), 10 (cyan dot-dashed curve)
and 15 (black dotted curve),
which are compared to the observations at the four corresponding redshifts,
shown as various symbols with corresponding colors.
The observational data are from \citet[][]{2015Bouwens}.
}
\label{fig:LF}
\end{figure}

\fi

We now address ``sources" of ionizing photons, in a fully self-consistent fashion,
in the standard cold dark matter model.
We follow the approach taken by \citet[][]{2015Trac},
to which the reader is referred for a more detailed description.
Briefly, the method uses direct observations of galaxy luminosity functions
at high redshift in the Hubble UDF to calibrate the star formation parameters
in the model based on halo mass accretion rate functions in the $\Lambda$CDM model.
Figure \ref{fig:LF} shows a comparison of rest-frame FUV luminosity functions
between the model based on the most recent cosmological parameters and observations 
at various redshifts.
The observed LFs are most reliable at $z\le 6$ and become less so towards higher redshifts,
and perhaps less than trustworthy beyond $z=8$ due to lack of spectroscopic confirmation at present.
For a given small region/area, such as the UDF, cosmic variance becomes more problematic towards higher redshift.
Additionally, it is possible that the observed LFs at high redshifts, in the midst of reionization,
may be masked by possible reionization effects;
this issue is significantly more acute for Ly$\alpha$ emitting galaxies
\citep[e.g.,][]{2004Mesinger,2005Haiman,2007Dijkstra}.
These problems can be circumvented, if we normalize the model at $z=6$ and use
the ``global" LFs from the model at high redshifts where direct observations lack or are unreliable.  
We take this approach.

From Figure \ref{fig:LF} we see that the model LFs  match observations well
at $z=6,7$. The agreement is still good at $z=8$, albeit with ``noisier" observational data.
There is very little to 
glean from the comparison at $z=10$, simply because the observational data lack both quantity and quality.
Integrating the Schechter fits of the \citet[][]{2015Bouwens} LF at $z=6$ yields
the intrinsic ionizing photon production rate from galaxies of
\begin{equation} \label{eq:Nionint}
\begin{split}
{\rm \dot N_{ion,int}} &={\rm 10^{51.52}cMpc^{-3}~s^{-1}\quad for\quad M_{UV,limit}=-12} \\
& = {\rm 10^{51.57}cMpc^{-3}~s^{-1}\quad for\quad M_{UV,limit}=-10} \\
& = {\rm 10^{51.61}cMpc^{-3}~s^{-1}\quad for\quad M_{UV,limit}=-8}.
\end{split}
\end{equation}
\noindent
In obtaining ${\rm \dot N_{ion,int}}$, 
we have used a relation between ionizing photo production rate per unit FUV spectral density
from \citep[][]{2013Robertson},
\begin{equation}
{\rm \xi_{ion} \equiv {{\dot N_{ion}/cMpc^{-3}~s^{-1}}\over {L_{UV}/erg~s^{-1}~Hz^{-1}~cMpc^{-3}}} = 10^{25.2}},
\label{eq:NionFUV}
\end{equation}
\noindent
which is based on the observed FUV spectral index $\beta\sim -2$ for high redshift galaxies.
Note $\beta$ is in defined in spectrum 
$f_\lambda d\lambda \propto \lambda^\beta d\lambda$, or $f_\nu d\nu \propto \nu^{-2-\beta}d\nu$,
in the FUV spectral range.
The accuracy of the normalization of our model
is such that the model LF at $z=6$ gives the same integrated light density 
as the observed one to the third digit.

Integrating the LF based on the $\Lambda$CDM model yield 
${\rm \dot N_{ion,int}(z=5.7) = 10^{51.6}cMpc^{-3}~s^{-1}}$,
weakly dependent on ${\rm M_{UV}}$ lower limit.
Dividing ${\rm \dot N_{ion,IGM}}$ in 
Eq \ref{eq:Nion} by ${\rm \dot N_{ion,int}(z=5.7)}$ gives
the mean luminosity-weighted escape fraction of Lyman continuum 
\begin{equation}
{\rm f_{esc,z=5.7} \equiv {\dot N_{ion,IGM}\over \dot N_{ion,int}} = 10\left({\dot N_{ion,IGM}\over 10^{50.6}cMpc^{-3}~s^{-1}}\right)\left({\xi_{ion}\over 10^{25.2}}\right)^{-1}\%}.
\label{eq:fesc}
\end{equation}
\noindent
We will show in \S 4 how ${\rm \dot N_{ion,IGM}}$ plays a key role in 
determining a lower bound on $\tau_e$ and how that in turn allow 
for a strong constraint on ${\rm \lambda_{mfp}}$ hence ${\rm M_{cut}}$.

\section{Reionization Histories Constrained by the State of IGM at $z=5.7$}

Any reionization history must satisfy the state of the IGM at $z=5.7$ and the fact that
the IGM is opaque to Ly$\alpha$ photon at just above that redshift. 
In this sense, the history of cosmological reionization becomes
a boundary value problem,
where we solve the evolution of HII volume fraction ${\rm Q_{HII}}$ with  
the following equation:
\begin{equation}
{\rm \frac{d Q_{HII}(z) }{dt} = \frac{\dot N_{ion,IGM}(z)}{n_{\rm H,0}} - \frac{Q_{HII}(z)}{t_{\rm rec}(z)}},
\label{eq:QHII}
\end{equation}
\noindent
where ${\rm n_{H,0}}$ is the comoving mean number hydrogen density, and \\
${\rm t_{rec}(z) = [C_{HII}(z)\,\alpha_{B}(T)~(1+Y_p/4[1-Y_p])~n_{H,0}~(1+z)^3]^{-1}}$ 
is the mean recombination time of ionized hydrogen in HII regions.
Any solution to Eq \ref{eq:QHII} satisfies the following two boundary conditions:
\begin{equation}
{\rm f_{esc}\dot N_{ion,int}\bar\sigma_{ion}\lambda_{mfp}|_{z=5.7} = {\dot N_{ion,IGM}\bar\sigma_{ion}\lambda_{mfp}|_{z=5.7} = 0.20_{-0.06}^{+0.11}\times 10^{-12} s^{-1}}}
\label{eq:fescNdot}
\end{equation}
\noindent
and 
\begin{equation}
{\rm Q_{HII}|_{z=5.7} = 1.0}.
\label{eq:QHIIz57}
\end{equation}
\noindent
In Eq \ref{eq:QHII} at $z>5.7$, since ${\rm \dot N_{ion,int}(z)}$ is fixed by the $\Lambda$CDM model
(see Figure \ref{fig:LF}),
we are left with only one degree of freedom, namely, the evolution of ${\rm f_{esc}}$ with redshift.
We model the redshift evolution 
of ${\rm f_{esc}}$ 
using a simple powerlaw form:
\begin{equation}
{\rm f_{esc}(z) = f_{esc,z=5.7} \left({1+z\over 6.7}\right)^\chi}.
\end{equation}
\label{eq:fescz}
Note that ${\rm f_{esc}(z)}$ in Eq \ref{eq:fescz}, like ${\rm f_{esc,z=5.7}}$ in Eq \ref{eq:fesc},  
is averaged over all the galaxies at a given redshift;
in other words, ${\rm f_{esc}(z)}$ is the ratio of the total number of ionizing photons entering 
the IGM to the total number of ionizing photons produced.
There is one additional physical process that is largely unconstrained by the state
of the IGM at $z=5.7$ but is important for the overall reionization history and integral electron scattering optical depth.
That is, a change of IMF at some high redshift from regular Pop II stars to a perhaps
more top-heavy and/or metal-free IMF,
which may lead to a quantitative transition in ionizing photon production efficiency per unit stellar mass,
${\rm \epsilon_{ion}}$.
Thanks to our lack of knowledge with regard to this process,
we choose to model ${\rm \epsilon_{ion}}$ generally, albeit in a simple way, as 
\begin{equation}
{\rm \epsilon_{ion} = \epsilon_{ion,PopII} + (\epsilon_{ion,PopIII}-\epsilon_{ion,PopII}) H(\Omega_{*}[z]-\Omega_{PopIII,crit})},
\end{equation}
\label{eq:eion}
\noindent
where 
${\rm \epsilon_{ion,PopIII}}$ and ${\rm \epsilon_{ion,PopII}}$ are 
ionizing photon production efficiency per unit stellar mass for Pop III and Pop II IMF, respectively.
We adopt ${\rm \epsilon_{ion,PopII}=3500\ photons/baryon}$ 
and \\
${\rm \epsilon_{ion,PopIII}=70000\ photons/baryon}$
\citep[e.g.,][]{2001bBromm},
resulting in ratio of \\
${\rm \epsilon_{ion,PopII}/\epsilon_{ion,PopIII}=20}$,
which enters our calculations.
The transition between Pop III and Pop II is modeled by a smoothed Heavyside step function
\begin{equation}
{\rm H(\Omega_{*}[z]-\Omega_{PopIII,crit}) = \left(1+\exp{[-2(\Omega_{*}(z)/\Omega_{PopIII,crit}-1)/\sigma_{PopIII}]}\right)^{-1}},
\end{equation}
\label{eq:H}
\noindent
where 
${\rm \Omega_{*}(z)}$ is the amount of stars formed by redshift ${\rm z}$ computed in the $\Lambda$CDM model in units of critical density,
${\rm \Omega_{PopIII,crit}}$, 
controls the transition from Pop III to Pop II when the amount of stars formed by some redshift in units of critical density
has reached this value, and ${\rm \sigma_{PopIII}}$ controls the width of this transition in units of 
${\rm \Omega_{PopIII,crit}}$; when ${\rm \sigma_{PopIII}=0}$, one recovers the unsmoothed Heavyside step function.
So far, we have three parameters to model the evolution of ionizing photon beyond $z=5.7$,
$\chi$, ${\rm \Omega_{PopIII,crit}}$ and ${\rm \sigma_{PopIII}}$. 
As we will show later, the dependence of results on 
${\rm \sigma_{PopIII}}$ 
is sufficiently weak that ${\rm \sigma_{PopIII}}$ can effectively be considered fixed,
as long as its value is not too large.
Therefore, we effectively have two free parameters in our model,
$\chi$ and ${\rm \Omega_{PopIII,crit}}$.
Given that we have one equation, Eq \ref{eq:QHII}, 
the general expectation is that there will be a family of solutions that will be able to meet
the two boundary conditions, Eq \ref{eq:fescNdot}, \ref{eq:QHIIz57}.
Conversely, though,
solving Eq \ref{eq:QHII} to obtain ${\rm Q_{HII}(z=5.7)=1}$ 
does not necessarily result in an IGM
at $z=5.7$ that is consistent with the constraint imposed 
by the observations of Ly$\alpha$ optical depth, i.e., 
Eq \ref{eq:fescz}, a point already noted by others \citep[e.g.,][]{2013Robertson}.




%

For each solution of ${\rm Q_{HII}(z)}$,
we compute the total electron scattering optical depth from $z=0$ to recombination redshift $z_{rec}$
by
\begin{equation}
{\rm \tau_e = \int_0^{z_{rec}}  f_e (1-f_s-f_n) Q_{H II} \sigma_T n_{H,0} [c/H(z)] (1+z)^{-1} dz},
\label{eq:taue}
\end{equation}
\noindent 
where ${\rm f_e}$ accounts for redshift evolution of helium contribution,
we use ${\rm f_e=(0.76+0.24/0.76/4)}$ for $z>2.8$ and 
${\rm f_e=(0.76+0.24/0.76/2)}$ for $z\le 2.8$, approximating
He II reionization as a step function at $z=2.8$,
which is consistent with 
the observed He II absorption optical depth data of \citep[][]{2011Worseck},
interpreted in the context of He II reionization simulations of \citep[][]{2009McQuinn}.
And ${\rm f_s}$ and ${\rm f_n}$ account for
stellar density and neutral hydrogen density, respectively,
which do not contribute to electron density.
\citet[][]{2008Wilkins} give $\Omega_*(z=0) = 2.5\times 10^{-3}$,
while \citep[][]{2015Grazian} yield $\Omega_*(z=6) = 3.7\times 10^{-5}$.
We interpolate between these two points to find an approximate stellar evolution fit
as $\Omega_*(z) = 2.5\times 10^{-3} (1+z)^{-2.1}$, translating to $f_s = 0.052 (1+z)^{-2.1}$. 
Post-reionization 
most of the neutral hydrogen 
resides in DLAs and the observational data
on the evolution of DLAs are available, albeit with significant errorbars.
We approximate the data presented in 
\citet[][]{2009Noterdaeme} by piece-wise powerlaws as follows:
$\Omega_{HI}=0.4\times 10^{-3}$ at $z=0$,
which evolves linearly to $\Omega_{HI}=0.9\times 10^{-3}$ at $z=0.5$,
which remains at $\Omega_{HI}=0.9\times 10^{-3}$ at $z=0.5-3$,
after which it linearly rises $\Omega_{HI}=1.2\times 10^{-3}$ at $z=3.5$,
followed by a constant $\Omega_{HI}=1.2\times 10^{-3}$ at $z=3.5-5.7$.

\ifmnras

\begin{figure}
\centering
\includegraphics[width=3.5in, keepaspectratio]{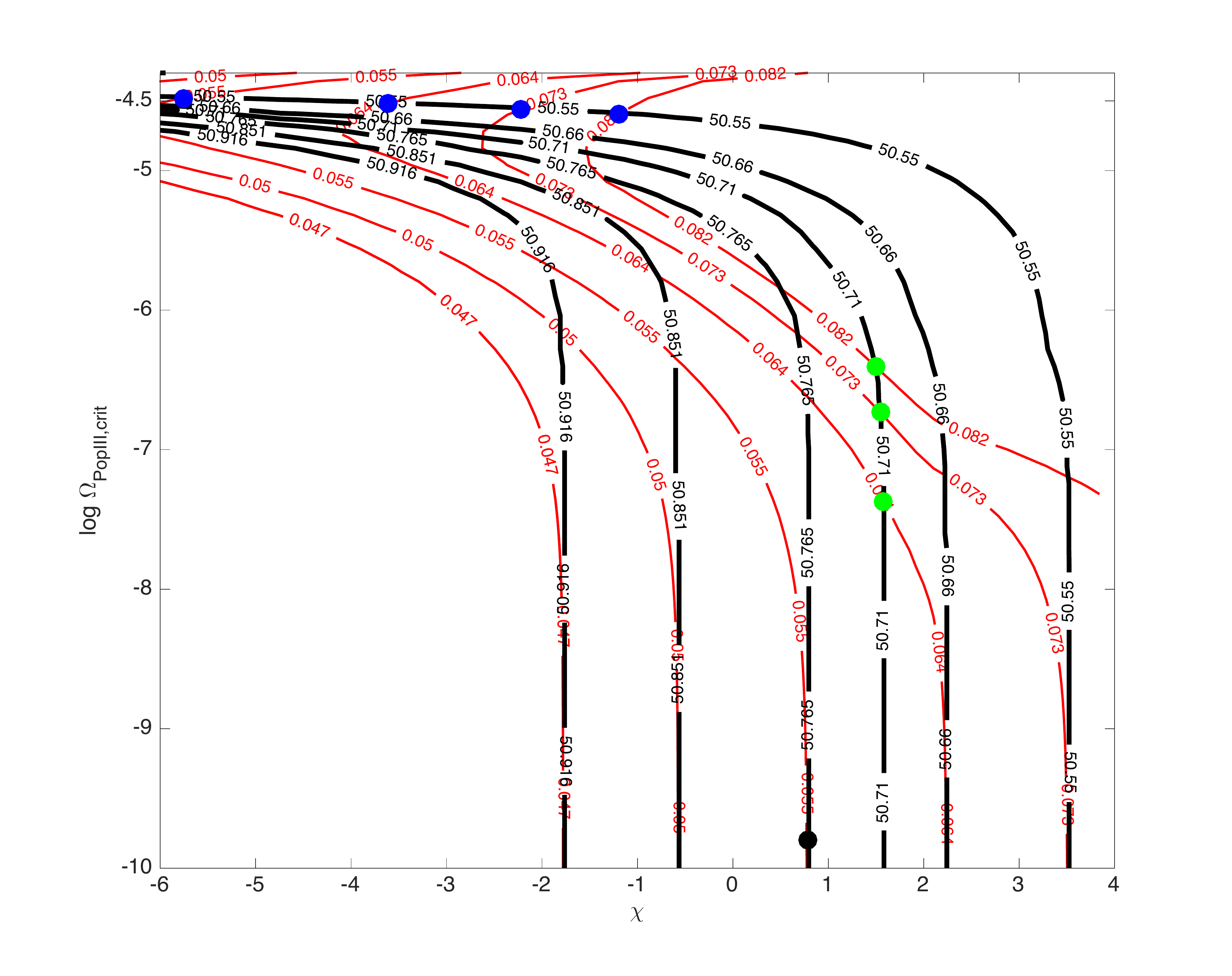}
\caption{
{\color{red}Left panel:}
shows the contours of $\tau_e$ (red) 
and ${\rm \dot N_{ion,IGM}(z=5.7)}$ (black)
in the $\chi-\Omega_{PopIII,crit}$ plane for ${\rm \sigma_{PopIII}=0.25}$.
The red contours are labelled with $\tau_e$ values,
whereas the black contours are labelled with 
${\rm \log \dot N_{ion,IGM}(z=5.7)}$ values.
The {\color{blue}four blue solid dots}
indicate four possible solutions of ${\rm Q_{HII}(z)}$
that yield total electron optical depths of ${\tau_e = (0.055,0.064,0.073,0.082)}$, respectively,
from left to right.
The {\color{green} three green solid dots}
indicate another set of three possible solutions of ${\rm Q_{HII}(z)}$
that yield total electron optical depths of ${\tau_e=(0.082,0.073,0.064)}$, respectively, from top to bottom.
The black solid dot is a solution with ${\tau_e=0.055}$. 
These specific solutions are discussed in the text.
}
\label{fig:zoom}
\end{figure}

\else

\begin{figure}[!h]
\begin{center}
\includegraphics[width=5.5in, keepaspectratio]{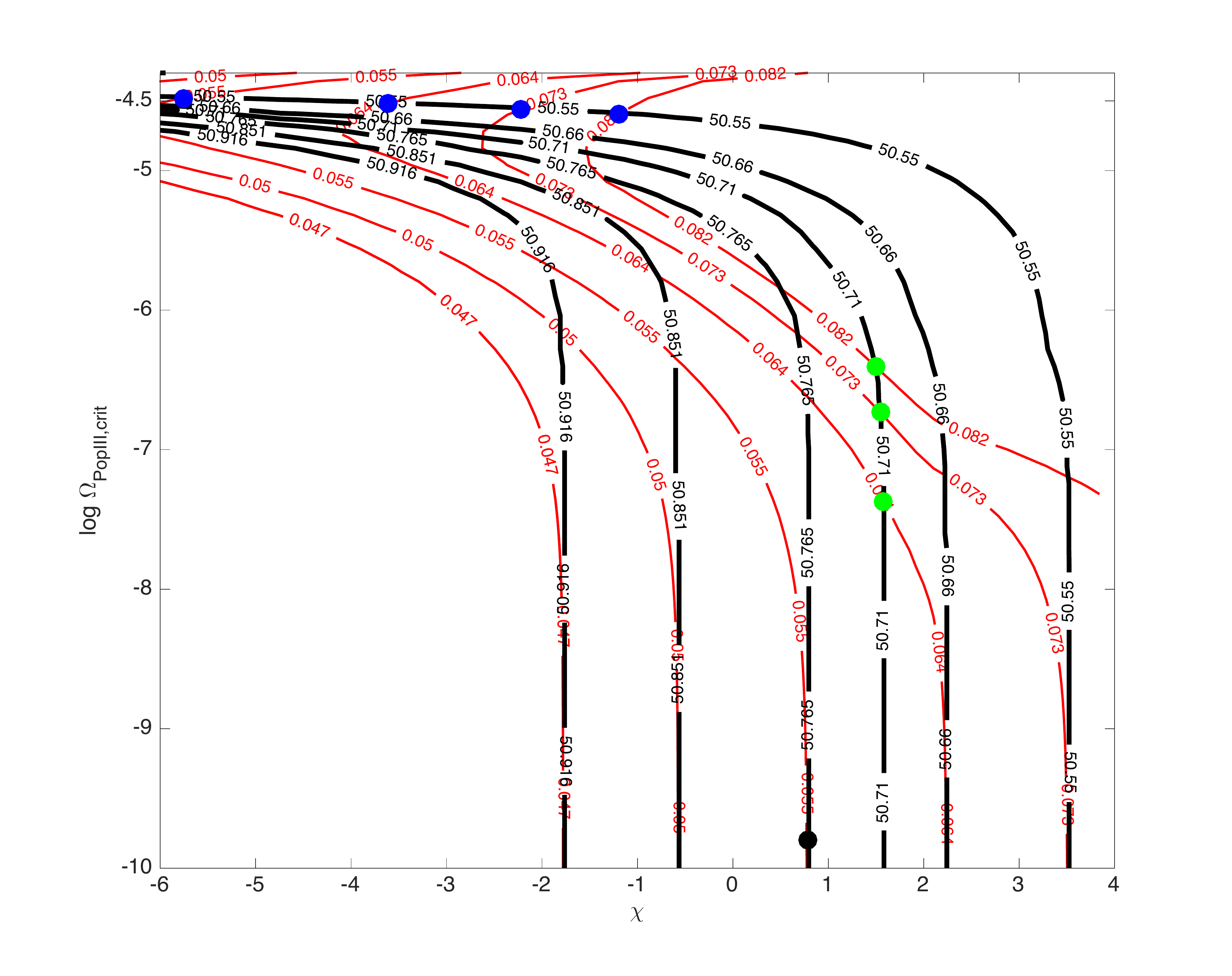}
\end{center}
\vskip -1.0cm
\caption{
shows the contours of $\tau_e$ (red) 
and ${\rm \dot N_{ion,IGM}(z=5.7)}$ (black)
in the $\chi-\Omega_{PopIII,crit}$ plane for ${\rm \sigma_{PopIII}=0.25}$.
The red contours are labelled with $\tau_e$ values,
whereas the black contours are labelled with 
${\rm \log \dot N_{ion,IGM}(z=5.7)}$ values.
The {\color{blue}four blue solid dots}
indicate four possible solutions of ${\rm Q_{HII}(z)}$
that yield total electron optical depths of ${\tau_e = (0.055,0.064,0.073,0.082)}$, respectively,
from left to right.
The {\color{green} three green solid dots}
indicate another set of three possible solutions of ${\rm Q_{HII}(z)}$
that yield total electron optical depths of ${\tau_e=(0.082,0.073,0.064)}$, respectively, from top to bottom.
The black solid dot is a solution with ${\tau_e=0.055}$. 
These specific solutions are discussed in the text.
}
\label{fig:zoom}
\end{figure}

\fi


Figure \ref{fig:zoom} shows the case with ${\rm \sigma_{PopIII}=0.25}$, to be examined in greater details.
We have examined cases with ${\rm \sigma_{PopIII}=0.5,0.25,0.05,0.01}$ and find that
the results, as displayed in Figure \ref{fig:zoom} in terms of the contours,
depend weakly on ${\rm \sigma_{PopIII}}$.
We note that the conclusions obtained are generic
and more importantly, the solution family obtained that is still viable is very insensitive to the choice of 
${\rm \sigma_{PopIII}}$.

It proves useful for our discussion to rewrite one of the boundary value constraints, namely, Eq \ref{eq:fescNdot}, 
as
\begin{equation}
{\rm \dot N_{ion,IGM}(z=5.7) = (1.8-4.1)\times 10^{50}\left[{\lambda_{mfp}(z=5.7)\over 7.6pMpc}\right]^{-1} cMpc^{-3}~s^{-1}},
\label{eq:NionIGM}
\end{equation}
\noindent
where the range inside the first pair of parentheses on the right hand side corresponds to $1\sigma$ lower and upper limits of Eq \ref{eq:gamma12}.
In this parameter space of ${\rm \chi-\Omega_{PopIII, crit}}$ shown in Figure \ref{fig:zoom} 
we have solutions to 
Eq \ref{eq:QHII} that satisfy Eq \ref{eq:QHIIz57}, i.e., the universal reionization completes exactly at $z=5.7$ 
with varying ${\rm \dot N_{ion,IGM}(z=5.7)}$ shown as the black contours.
Superimposed as the red contours are values of ${\rm \tau_e}$ for each solution.

\ifmnras

\begin{figure}
\centering
\includegraphics[width=3.5in, keepaspectratio]{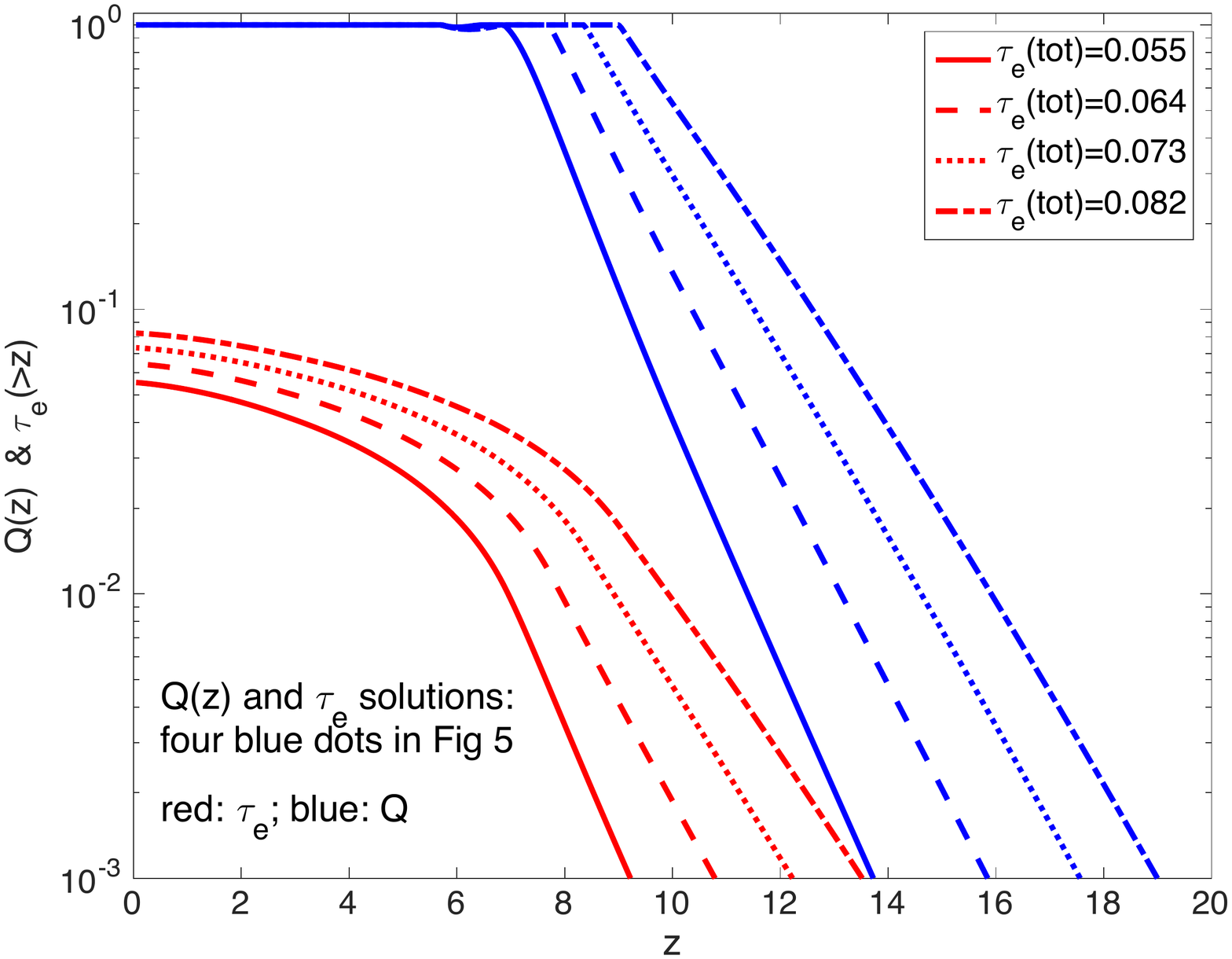}
\caption{
{\color{red}Left panel:}
shows each of the four solutions of ${\rm Q_{HII}(z)}$ (blue curves) indicated by 
the {\color{blue}four blue solid dots} in Figure \ref{fig:zoom},
along with the respective cumulative ${\rm \tau_e}$ (red curves).
}
\label{fig:bb}
\end{figure}

\else

\begin{figure}[!h]
\begin{center}
\includegraphics[width=5.5in, keepaspectratio]{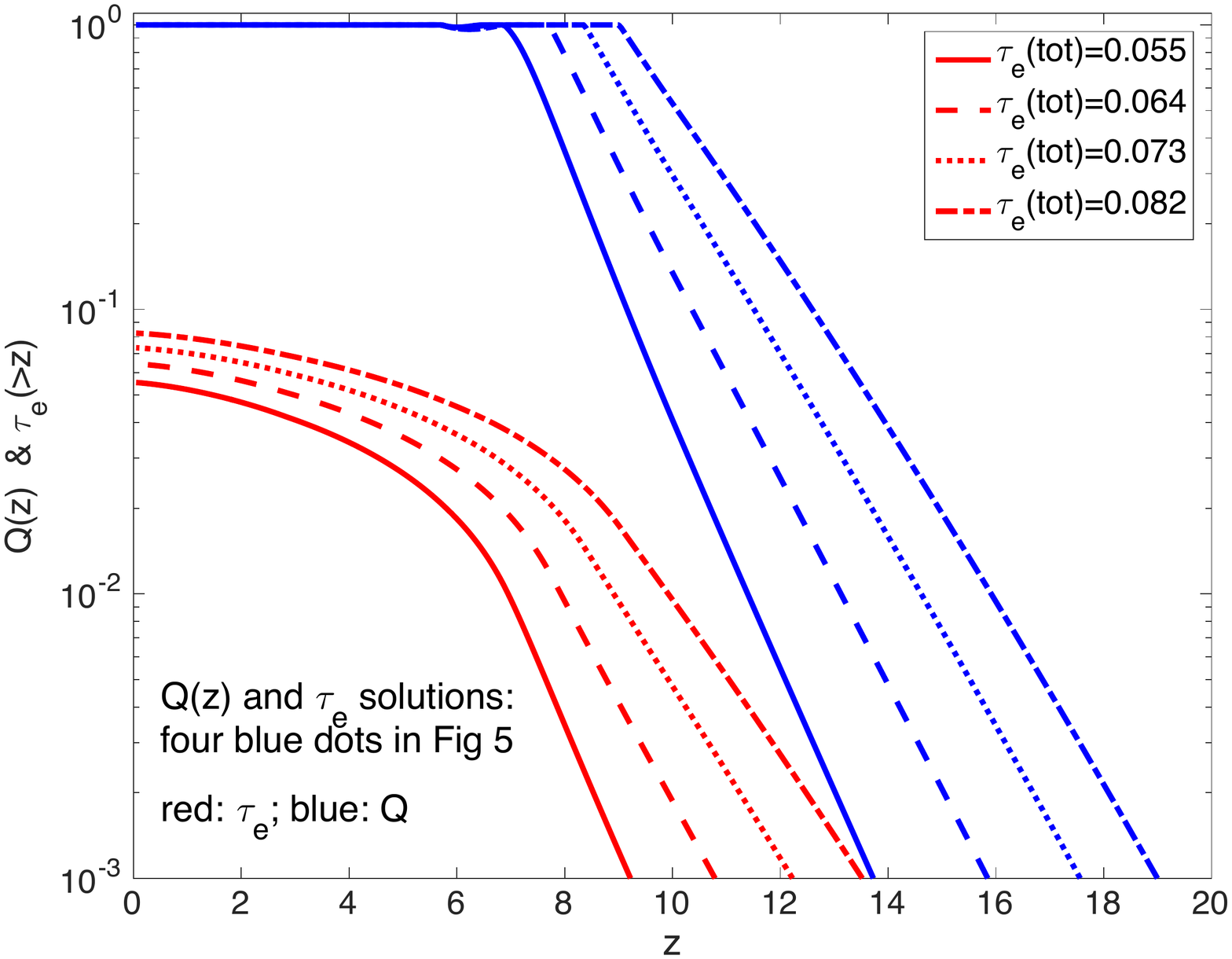}
\end{center}
\vskip -1.0cm
\caption{
shows each of the four solutions of ${\rm Q_{HII}(z)}$ (blue curves) indicated by 
the {\color{blue}four blue solid dots} in Figure \ref{fig:zoom},
along with the respective cumulative ${\rm \tau_e}$ (red curves).
}
\label{fig:blue}
\end{figure}

\fi

It is now clear that the value of ${\rm \dot N_{ion,IGM}(z=5.7)}$ plays 
a key role in determining the viability of each solution of ${\rm Q_{HII}(z)}$.
Under the two boundary conditions, Eq \ref{eq:fescNdot} and \ref{eq:QHIIz57},
two families of solutions are possible, each of which is {\it simultaneously} 
consistent with the latest values of ${\rm \tau_e}$ from \citet[][]{2016Planck} observations.
Indicated by the {\color{blue} four blue dots}
in Figure \ref{fig:zoom} 
are four solutions in the (we call) ``Pop III-supported" family with 
${\rm \tau_e=(0.055,0.064,0.073,0.082)}$ corresponding 
to the (central, $+1\sigma$, $+2\sigma$, $+3\sigma$) values from \citet[][]{2016Planck}.

Figure \ref{fig:blue} 
shows each of the four solutions of ${\rm Q_{HII}(z)}$ (blue curves) indicated by 
the {\color{blue}four blue solid dots} in Figure \ref{fig:zoom},
along with the respective cumulative ${\rm \tau_e}$ (red curves).
The common characteristics of these solutions in this solution family are that
(1) $\chi<0$, indicating that the escape fraction decreases with increasing redshift,
(2) the Pop III stars make a significant and late contribution to the overall ionizing photon budget.
The combination of negative $\chi$ and late, significant Pop III contribution
permits a slight dip in ionized fraction at a redshift slightly higher than $z=5.7$,
to satisfy \ref{eq:QHIIz57}.
This set of solutions, however, may be inconsistent with some other independent observations.
Here we provide some notable examples.

\ifmnras

\begin{figure}
\centering
\includegraphics[width=3.5in, keepaspectratio]{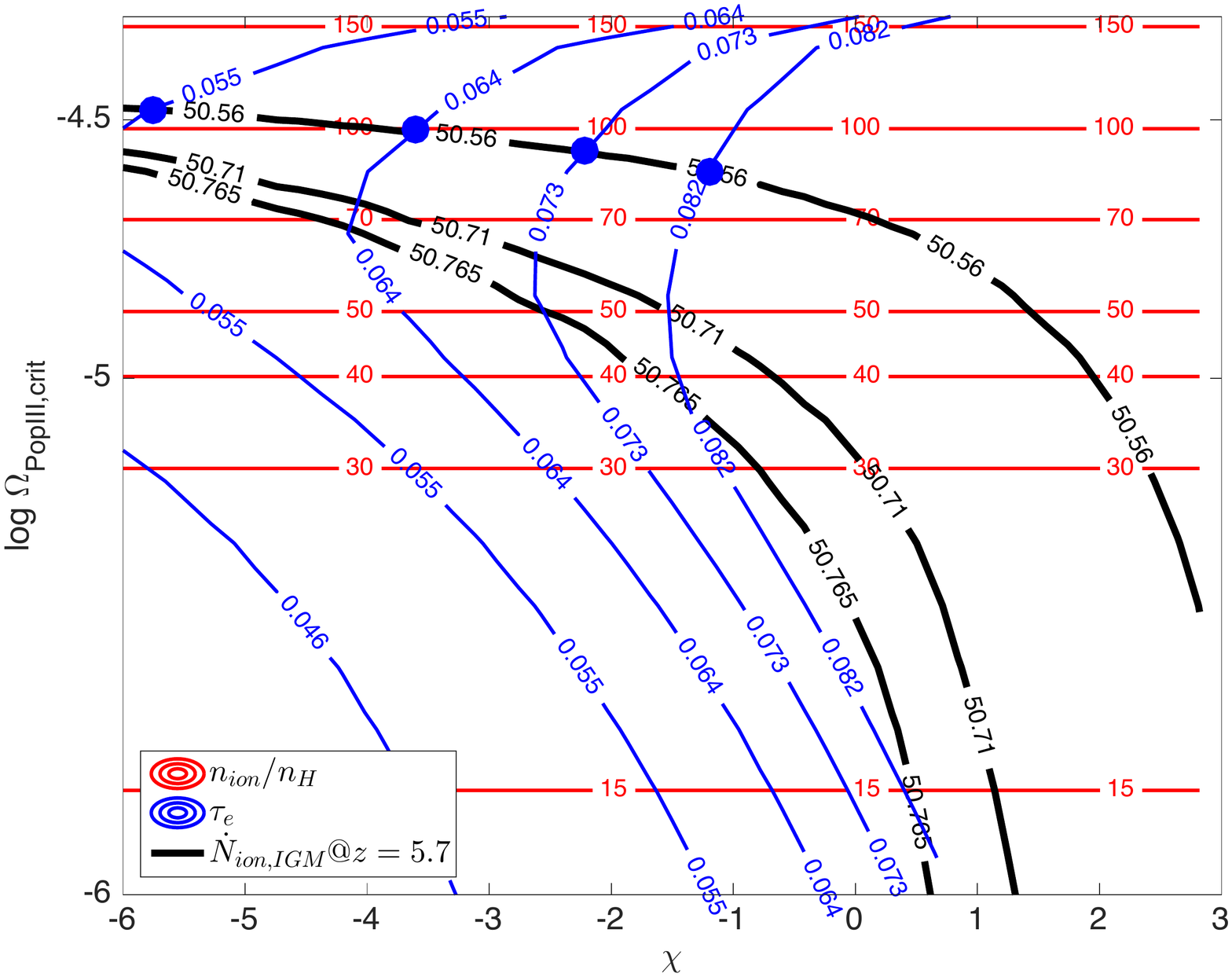}
\caption{
{\color{red}Left panel:}
shows contours of the ratio of the number of ionizing photon produced per hydrogen atom (red),
along with contours of ${\rm \tau_e}$ (blue) and 
of ${\rm \log \dot N_{ion,IGM}(z=5.7)}$ (black).
}
\label{fig:np}
\end{figure}

\else

\begin{figure}[!h]
\begin{center}
\includegraphics[width=5.5in, keepaspectratio]{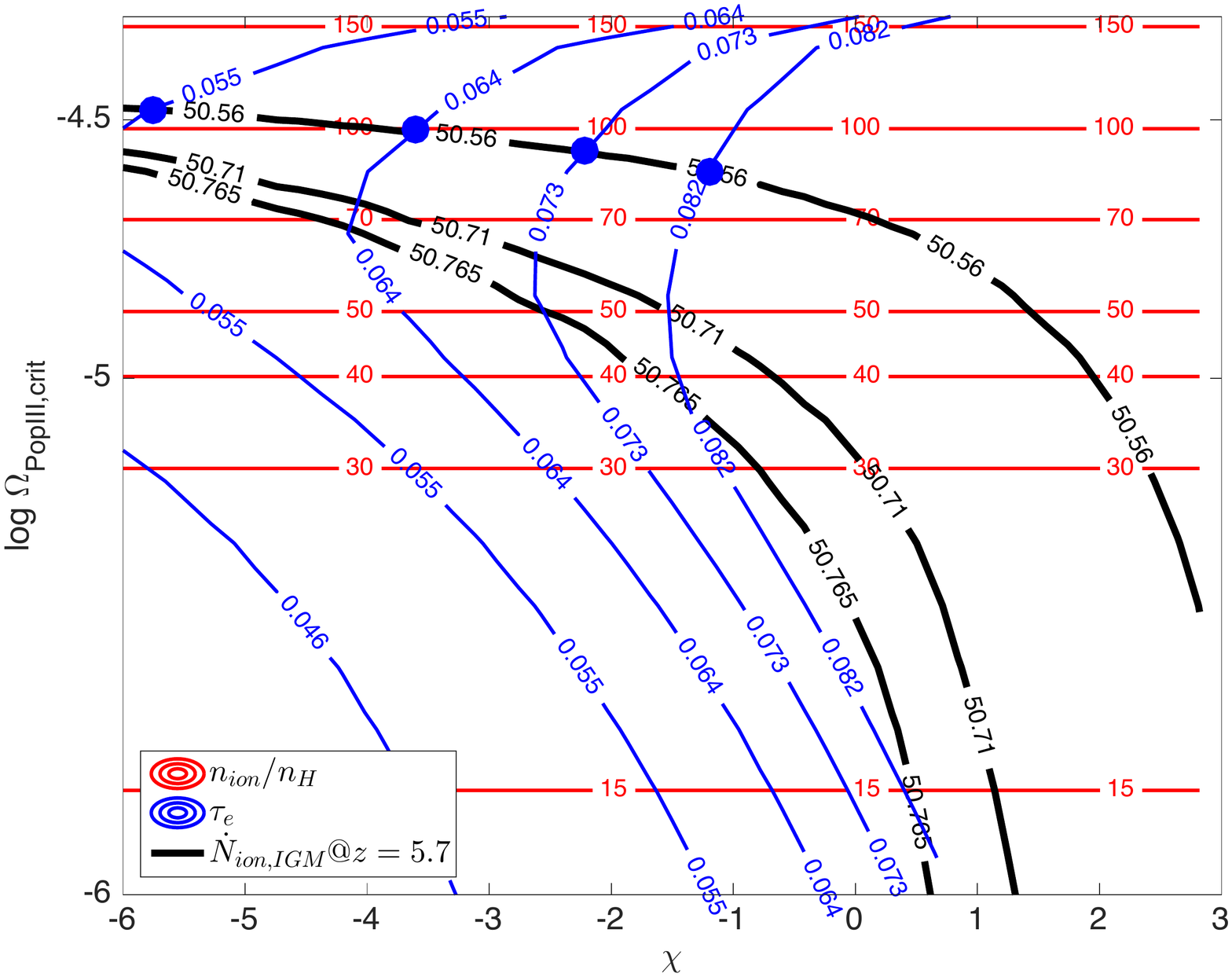}
\end{center}
\vskip -1.0cm
\caption{
shows contours of the ratio of the number of ionizing photon produced per hydrogen atom (red),
along with contours of ${\rm \tau_e}$ (blue) and 
of ${\rm \log \dot N_{ion,IGM}(z=5.7)}$ (black).
}
\label{fig:np}
\end{figure}

\fi

Figure \ref{fig:np} 
shows contours of the ratio of the number ionizing photon produced per hydrogen atom (red).
\citet[][]{2004Fang} perform a detailed analysis of metal enrichment history 
and show that Pop III to Pop II transition occurs when $3-20$ ionizing photons per hydrogen atom,
depending on the model for the IMF,
have been produced by Pop III stars, based on considerations of primary atomic cooling agents, CII and OI,
at low temperature, corresponding to ${\rm [C/H]_{crit}=-3.5}$ and ${\rm [O/H]_{crit}=-3.1}$
\citep[][]{2003bBromm}.
For the four solutions, indicated by the four blue dots in Figure \ref{fig:np},
we see much higher, $80-110$ ionizing photons per hydrogen atom have been produced at the 
model transition ${\rm \Omega_{PopIII, crit}}$, 
in order to attain the solutions.
Note that in the scenario of dust cooling induced fragmentation 
\citep[][]{2010Schneider}, the critical transition metallicity 
is $1-3$ orders of magnitude lower that is still more stringent. 
These considerations indicate that these ${\rm Q_{HII}(z)}$ solutions are self-inconsistent,
in the sense that the required Pop III contribution in order for the solutions to be possible 
is unattainable.

A second example concerns the neutral fraction of the IGM 
during the epoch of reionization at $z>6$.
In a recent careful analysis of possible signatures of damping wing absorption profiles
of the Ly$\alpha$ emission line of quasar J1120+0641 at $z=7.1$,
under the assumption that DLAs, being sufficiently rare, are not responsible for the absorption
of the Ly$\alpha$ emission redward of the line,
\citet[][]{2016Greig} conclude that the mean neutral fraction of the IGM
is $0.40^{+0.41}_{-0.32}$ ($2\sigma$).
All of the four solutions shown in Figure \ref{fig:blue} have the mean neutral fraction significantly 
less than a few percent, thus are ruled out at $>2.5\sigma$ level.

\ifmnras

\begin{figure}
\centering
\includegraphics[width=3.5in, keepaspectratio]{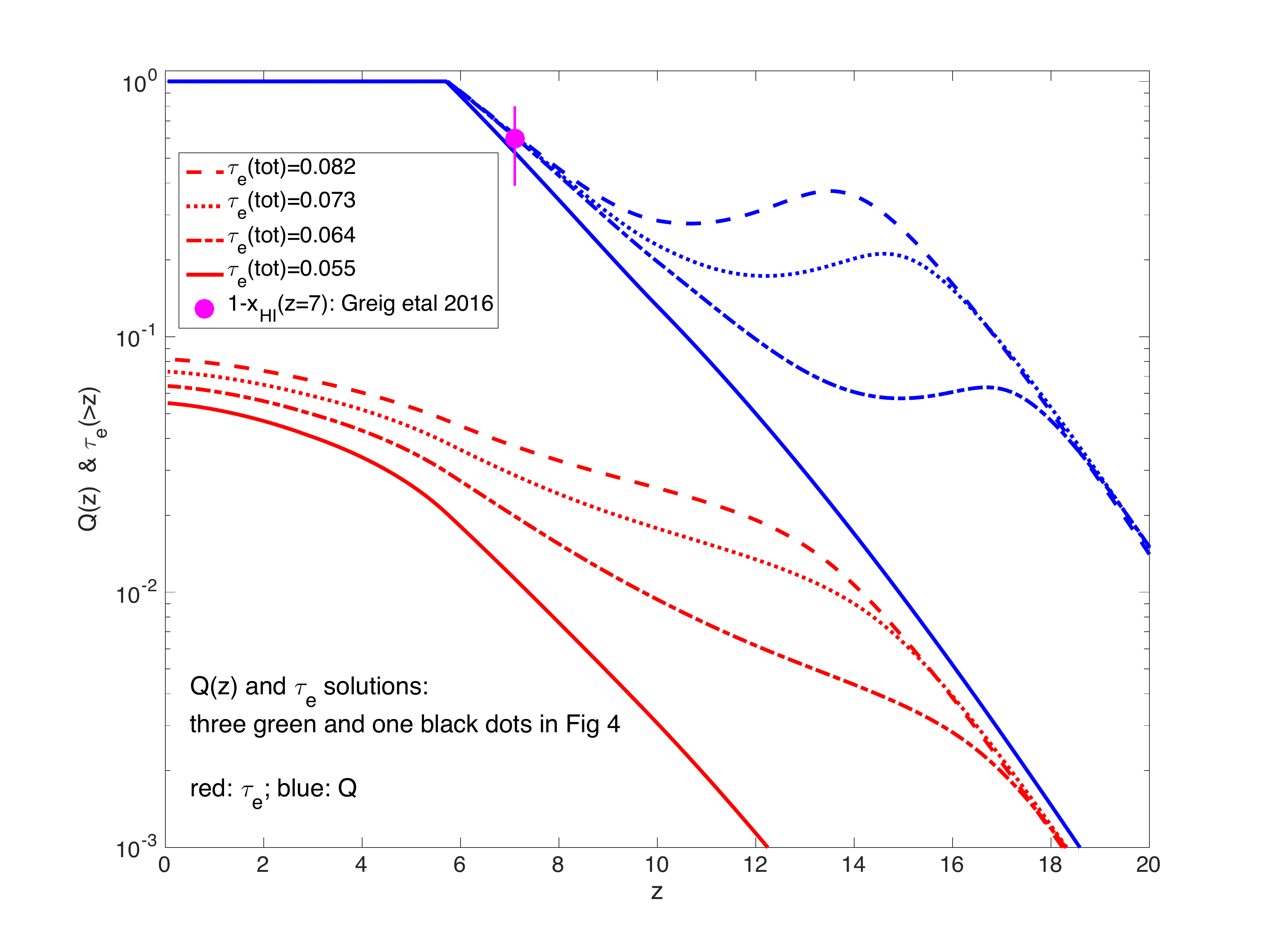}
\caption{
{\color{red}Left panel:}
shows each of the four solutions of ${\rm Q_{HII}(z)}$ (blue curves) indicated by 
the {\color{green}three green} (for ${\rm \tau_e=(0.064,0.0740.082)}$)
and 
one black (for ${\rm \tau_e=0.055}$) solid dots in Figure \ref{fig:zoom},
along with the respective cumulative ${\rm \tau_e(>z)}$ (red curves).
Indicated by the magenta solid dot is an observational measurement
of neutral fraction of the IGM at $z=7.1$ by 
\citet[][]{2016Greig} based on the damping wing signature imprinted
on the red side of the Ly$\alpha$ emission line of quasar J1120+0641.
}
\label{fig:green}
\end{figure}

\else

\begin{figure}[!h]
\begin{center}
\includegraphics[width=5.5in, keepaspectratio]{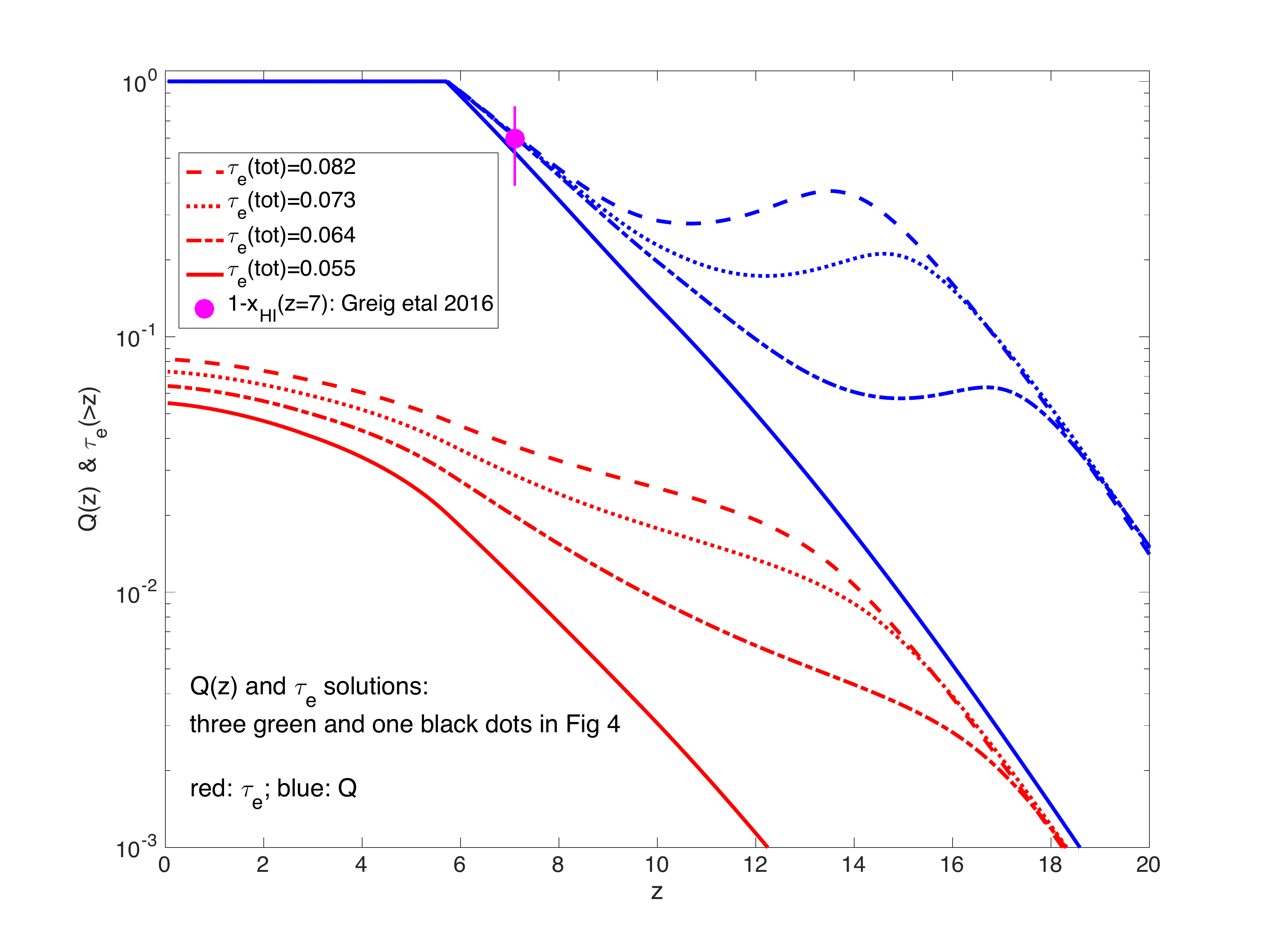}
\end{center}
\vskip -1.0cm
\caption{
shows each of the four solutions of ${\rm Q_{HII}(z)}$ (blue curves) indicated by 
the {\color{green}three green} (for ${\rm \tau_e=(0.064,0.0740.082)}$)
and 
one black (for ${\rm \tau_e=0.055}$) solid dots in Figure \ref{fig:zoom},
along with the respective cumulative ${\rm \tau_e(>z)}$ (red curves).
Indicated by the magenta solid dot is an observational measurement
of neutral fraction of the IGM at $z=7.1$ by 
\citet[][]{2016Greig} based on the damping wing signature imprinted
on the red side of the Ly$\alpha$ emission line of quasar J1120+0641.
}
\label{fig:green}
\end{figure}

\fi

Let us now turn to the other solution family with 
reduced Pop III contribution that is additionally confined to much higher redshift.
Figure \ref{fig:green} 
shows each of the three solutions of ${\rm Q_{HII}(z)}$ (blue curves) indicated by 
the {\color{green}three green} and one solid dots in Figure \ref{fig:zoom},
along with the respective cumulative ${\rm \tau_e}$ (red curves).
Several trends shared by solutions in this solution family may be noted.
First, ${\rm Q_{HII}(z)}$ increases exponentially as a function of redshift
in the range of $z=5.7$ to $z=9-14$, depending on the value of total ${\rm \tau_e}$;
a lower total ${\rm \tau_e}$ corresponds to a higher redshift, but lower value of 
${\rm Q_{HII}(z)}$ base, from which the exponential growth starts.
All four solutions are consistent with the observationally inferred
mean neutral fraction of the IGM at $z=7.1$,
shown as a magenta dot with $1\sigma$ range
\citep[][]{2016Greig}.
Second, there is a distinct, separate peak ${\rm Q_{HII}(z)}$ at 
$z=14-18$, for ${\rm \tau_e=0.082-0.064}$ (in that order) with height of 
$(0.4-0.07)$ (in the same order).
This high redshift peak of ${\rm Q_{HII}(z)}$ is due to contributions from Pop III stars.
The exact height and duration of this peak may depend on the assumptions
concerning the transition from Pop III to Pop II temporally and spatially,
that will require detailed modeling beyond the scope of this work. 
We note, however, that the results do not change significantly
when values of ${\rm \sigma_{III}=0.01-1}$ are used ($0.25$ is used for the case shown in 
Figure \ref{fig:green}),
suggesting that the existence, the ${\rm Q_{HII}(z)}$ value of the peak and the peak redshift are fairly robust. 
We also note that all these solutions lie below ${\rm \Omega_{PopIII, crit}=10^{-6.4}}$, which, when compared 
with Figure \ref{fig:np}, indicates a consistency in terms of 
Pop III stars forming in the metallicity regime that is physically plausible,
if low temperature atomic cooling, not dust cooling, dictates fragmentation of star-forming gas clouds. 
Finally, it is seen that these solutions have $\chi\ge 0$, 
indicating that the escape fraction increases with increasing redshift,
perhaps not an unexpected result based on physical considerations that galaxies at high redshifts
are less massive, their star-formation episodes more bursty and consequently their interstellar medium more 
porous
to allow for more ionizing photons to escape.
Simulation results are consistent with this trend \citep[e.g.,][]{2014Kimm}.
In summary, this solution family are self-consistent.

If, however, ${\rm\tau_e=0.055}$ holds up,
there is no solution of ${\rm Q_{HII}(z)}$ with \\
${\rm \log \dot N_{ion,IGM}(z=5.7)=50.71}$.
In order to get a solution with ${\rm\tau_e=0.055}$,
one requires \\ 
${\rm \log \dot N_{ion,IGM}(z=5.7)=50.765}$,
which, with the conservative choice of $+1\sigma$ value ${\rm \Gamma_{-12}=0.31}$ (see Eq \ref{eq:gamma12}),
in turn requires ${\rm \lambda_{mfp}(z=5.7)=5.3pMpc}$, 
which would be at about $2.9\sigma$ lower bound of the observationally inferred value.
In combination with the $+1\sigma$ value of $\Gamma_{12}$ used,
such an event would be a $3.0\sigma$ occurrence,
suggesting tension,
which we examine in the next section.

\section{${\rm \lambda_{mfp}(z=5.7)}$: A Strong Test of Matter Power Spectrum on Small Scales}

We were left in a state of significant tension between accommodating ${\rm \tau_e=0.055}$ 
and ${\rm \lambda_{mfp}(z=5.7)}$ based on the extrapolated observational data at $z<5.5$ in \S 3.
The tension may be alleviated, if one chooses not to strongly advocate the central value
of ${\rm \tau_e=0.055}$ \citep[][]{2016Planck} but instead emphasize the harmonious 
concordance between ${\rm \lambda_{mfp}(z=5.7)}$, $\Gamma(z=5.7)$ and ${\tau_e\ge 0.64}$.
We take this discrepancy in a somewhat different way and 
suggest that the extrapolation of the lower redshift
measurement of ${\rm \lambda_{mfp}}$ should be taken with caution,
despite the smooth trend seen in the observed redshift range ($z=2.3-5.5$).
We take a step further yet to perform a theoretical analysis to better understand the
physical origin of ${\rm \lambda_{mfp}(z=5.7)}$ in the context of the standard cosmological model.

It is useful to separate out the overall 
${\rm \lambda_{mfp}}$ into two components in the post-overlap epoch at $z=5.7$,
one due to the ``translucent", general volume-filling low density IGM that collectively 
attenuates ionizing photons
and the other due to "opaque" disks (like LLSs) that block entirely all incident ionizing photons.
We shall denote them ${\rm \lambda_{mfp,IGM}}$ and ${\rm \lambda_{mfp,halo}}$, respectively.
The total ${\rm \lambda_{mfp}}$ is
\begin{equation} 
{\rm \lambda_{mfp} = (\lambda_{mfp,halo}^{-1} + \lambda_{mfp,IGM}^{-1})^{-1}}.
\label{eq:lambda}
\end{equation}
The ${\rm \lambda_{mfp,IGM}}$ can be approximated by 
the volume-weighted neutral fraction of the IGM as
\begin{equation}
\begin{split}
{\rm \lambda_{\rm mfp, IGM}}& = {\rm (\bar\sigma_{ion} f_{HI,vol} n_{H,0}(1+z)^3)^{-1}}\\
&={\rm 19.5 ({1+z\over 6.7})^{-3} ({\bar\sigma_{ion}\over 3.16\times 10^{-18} cm^{2}})^{-1} ({f_{HI,vol}\over 0.9\times 10^{-4}})^{-1}~pMpc},
\end{split}
\label{eq:lambdaIGM}
\end{equation}
\noindent
where ${\rm f_{HI,vol}=0.9\times 10^{-4}}$ is the volume-weighted neutral fraction of the IGM,
inferred by the directly observed Ly$\alpha$ (and higher order Lyman transitions) optical depth
at $z=5.7$ \citep[][]{2006Fan}. 
As we have argued earlier, 
while the mass-weighted neutral fraction determined from such a method
may be significantly model-dependent, the volume-weighted neutral fraction is not expected to be, because
it is free from clumping factor dependence and 
most of the optical depth contributions stem from low-density regions of optical depth of order unity
whose Jeans scales are typically resolved in most simulations used.

${\rm \lambda_{mfp,halo}}$ stems from self-shielding dense gas in halos.
A computation of ${\rm \lambda_{mfp,halo}}$ may not seem a well posed problem at first sight,
because it would appear to depend on both the abundance of halos and their cross sections (the sizes of 
radiation blocking disks).
It is not immediately obvious how one may precisely specifiy their cross sections,
even if their abundance is known.
We show that this ambiguity can be removed, 
when considerations are given to the physical conditions of halo gas as a function of halo-centric radius and
a ``correct" definition of ${\rm \lambda_{mfp,halo}}$ is adopted,
which we now describe.

After the HII regions have overlapped in the aftermath of reionization,
neutral gas in halos essentially becomes a set of disconnected isolated
islands that are increasingly self-shielded and optically thick to ionizing photons toward to the centers of halos.
Under the assumption of spherical symmetry, for a given halo,
we can compute the column density as a function of halo-centric radius $r$ outside-in as
\begin{equation}
{\rm N_{HI}(r) = \int_r^\infty x_{HI}(r^\prime)\delta(r^\prime)n_{H,0}(1+z)^3dr^\prime},
\label{eq:NHIr}
\end{equation}
\noindent
where $\delta(r)\equiv n(r)/\bar n$ is overdensity, for which we use
the universal halo density profile \citep[NFW, ][]{1997Navarro} 
with gas following mass over the relevant radial range \citep[e.g.,][]{2001Komatsu}.
In the core region of a halo the gas density is constrained such that
the gas entropy does not fall below the entropy of the gas at the mean density
and cosmic microwave background temperature.
In practice, the upper limit of the integral in Eq \ref{eq:NHIr} is chosen
when $\delta=1$ (i.e., the mean density) but its precise value makes no material difference to 
the calculated ${\rm N_{HI}(r)}$ in the range of relevance.
The local neutral fraction ${\rm x_{HI}(r)}$ at radius $r$ 
can be computed using the local balance between recombination and photoionization
through a spherical radiative transfer:
\begin{equation}
{\rm \Gamma \exp{[-N_{HI}(r)\bar\sigma_{ion}]} x_{HI}(r) = [1-x_{HI}(r)]^2 [1+Y_p/4(1-Y_p)] \alpha_B(T)\delta(r) n_{H,0}(1+z)^3},
\label{eq:local}
\end{equation}
\noindent
where $\Gamma$ is the ``background" ionization rate prior to significant attenuation when approaching
the halo.
We solve Eq (\ref{eq:NHIr},\ref{eq:local}) numerically to obtain ${\rm N_{HI}(r)}$ and
${\rm x_{HI}(r)}$, for a given $\Gamma$.

\ifmnras

\begin{figure}
\centering
\includegraphics[width=3.5in, keepaspectratio]{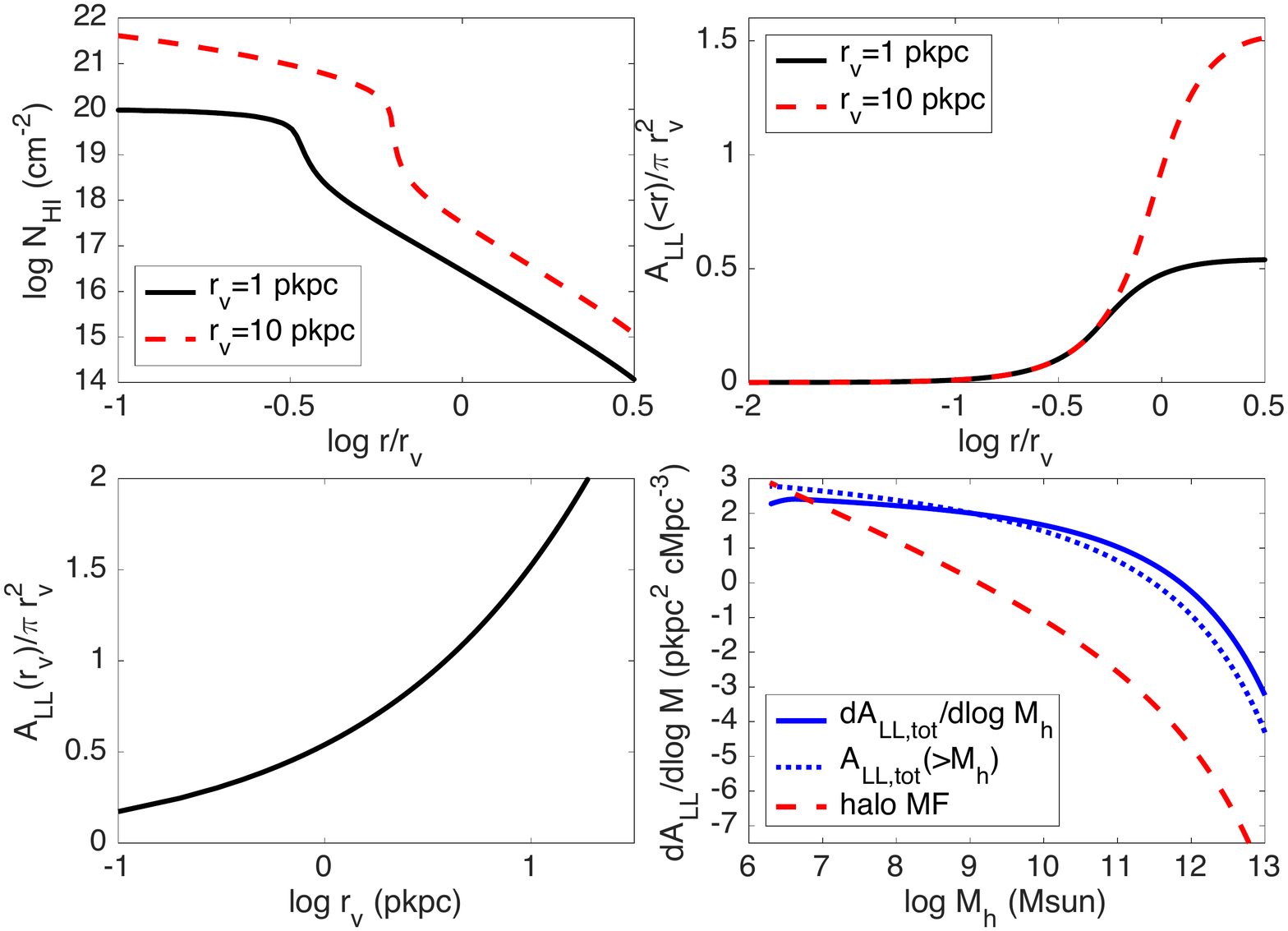}
\caption{
shows the integrated 
column density (from outside inward down to the halo-centric radius ${\rm r}$) 
as a function of ${\rm r}$ (in units of virial radius ${\rm r_v}$),
for two cases with virial radius ${\rm r_v}$ equal to $1~$pkpc (black solid curve)
and $10~$pkpc (red dashed curve) at $z=5.7$
[with corresponding virial (temperature, mass) of 
(${\rm 1.5\times 10^3}$K, ${9.7\times 10^6\msun}$)
and (${\rm 1.5\times 10^5}$K, ${9.7\times 10^9\msun}$), respectively],
using fiducial values for various parameters:
$\Gamma_{-12}=0.2$, ${\rm\bar\sigma_{ion}=3.16\times 10^{-18}~cm^{2}}$.
For the NFW profile we use a concentration parameter ${\rm C=5}$ in both cases.
{\color{red}Top-right panel:}
shows the cumulative cross section for ionizing photons of a halo 
${\rm A_{LL}(<r_p)}$ in units of the virial area (${\rm\pi r_v^2}$)
as a function of halo-centric radius in units of the virial radius ${\rm r_v}$
for the two halos shown in the top-left panel of Figure \ref{fig:NFWcol}.
{\color{red}Bottom-left panel:}
shown the effective total cross section for Lyman continuum photons in units of halo virial area
as a function of halo virial radius at ${\rm z=5.7}$.
{\color{red}Bottom-right panel:}
shows the differential function ${\rm dA_{LL,tot}/d\log M_h\equiv {\rm n(M_h) M \ln 10 A_{LL}(M_h)}}$
as a function of ${\rm M_h}$ (solid blue curve), 
its cumulative function ${\rm A_{LL,tot}(>M_h)}$ (dotted blue curve),
along with the halo mass function ${\rm n(M_h) M \ln 10}$ as a function of 
${\rm M_h}$ (dashed red curve).
}
\label{fig:bb}
\end{figure}

\else

\begin{figure}[!h]
\begin{center}
\includegraphics[width=6.5in, keepaspectratio]{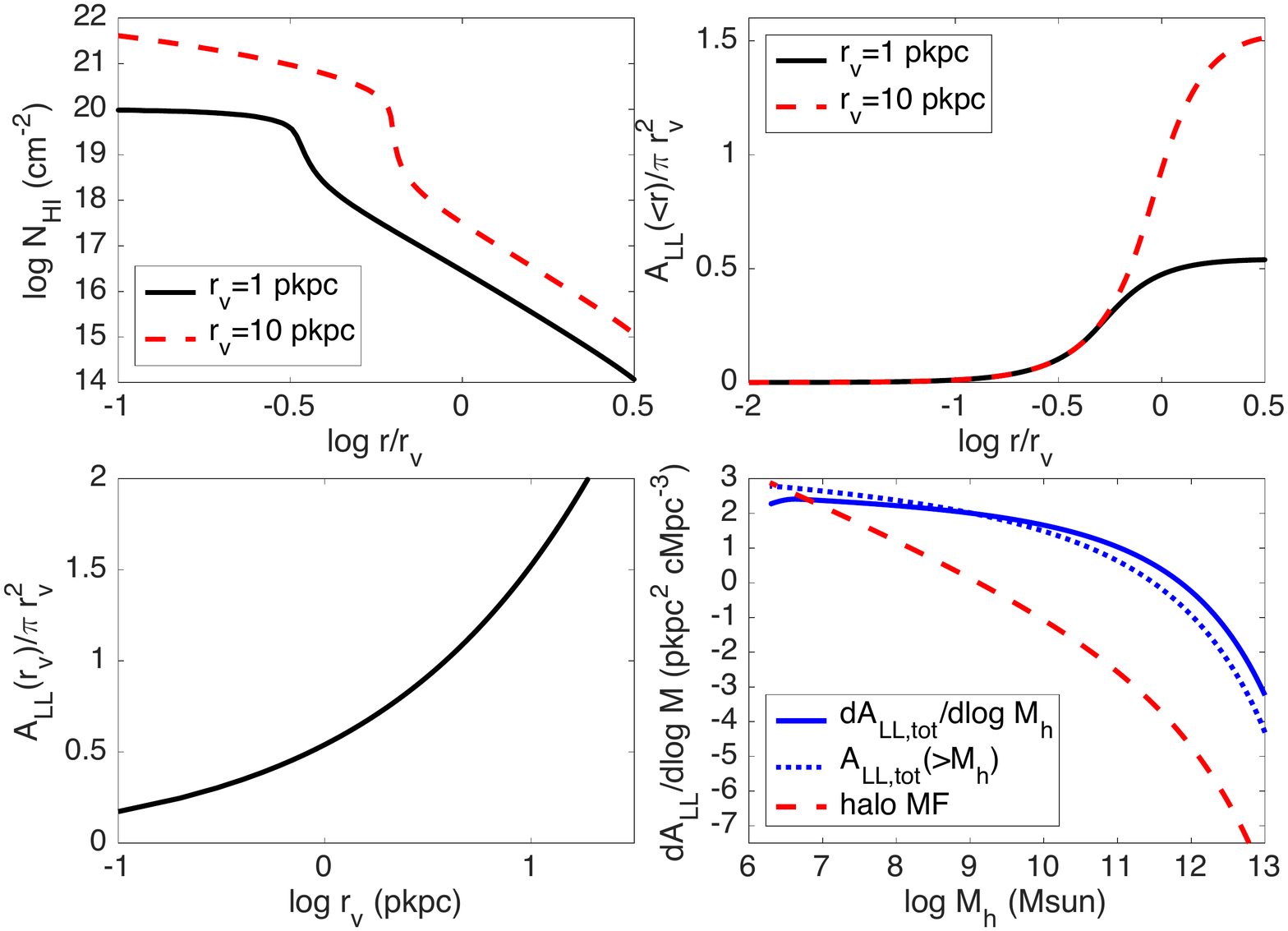}
\end{center}
\vskip -1.0cm
\caption{
{\color{red}Top-left panel:}
shows the integrated 
column density (from outside inward down to the halo-centric radius ${\rm r}$) 
as a function of ${\rm r}$ (in units of virial radius ${\rm r_v}$),
for two cases with virial radius ${\rm r_v}$ equal to $1~$pkpc (black solid curve)
and $10~$pkpc (red dashed curve) at $z=5.7$
[with corresponding virial (temperature, mass) of 
(${\rm 1.5\times 10^3}$K, ${9.7\times 10^6\msun}$)
and (${\rm 1.5\times 10^5}$K, ${9.7\times 10^9\msun}$), respectively],
using fiducial values for various parameters:
$\Gamma_{-12}=0.2$, ${\rm\bar\sigma_{ion}=3.16\times 10^{-18}~cm^{2}}$.
For the NFW profile we use a concentration parameter ${\rm C=5}$ in both cases.
{\color{red}Top-right panel:}
shows the cumulative cross section for ionizing photons of a halo 
${\rm A_{LL}(<r_p)}$ in units of the virial area (${\rm\pi r_v^2}$)
as a function of halo-centric radius in units of the virial radius ${\rm r_v}$
for the two halos shown in the top-left panel of Figure \ref{fig:NFWcol}.
{\color{red}Bottom-left panel:}
shown the effective total cross section for Lyman continuum photons in units of halo virial area
as a function of halo virial radius at ${\rm z=5.7}$.
{\color{red}Bottom-right panel:}
shows the differential function ${\rm dA_{LL,tot}/d\log M_h\equiv {\rm n(M_h) M \ln 10 A_{LL}(M_h)}}$
as a function of ${\rm M_h}$ (solid blue curve), 
its cumulative function ${\rm A_{LL,tot}(>M_h)}$ (dotted blue curve),
along with the halo mass function ${\rm n(M_h) M \ln 10}$ as a function of 
${\rm M_h}$ (dashed red curve).
}
\label{fig:NFWcol}
\end{figure}

\fi

In the top-left panel of 
Figure \ref{fig:NFWcol} we show the integrated 
column density (from outside inward down to the radius $r$) as a function of halo-centric radius $r$ (in units of virial radius ${\rm r_v}$)
for two cases with virial radius ${\rm r_v}$ equal to $1~$pkpc (black solid curve)
and $10~$pkpc (red dashed curve), respectively.
We see that at about ${\rm r/r_v\sim 3}$
the column density is well below ${\rm 10^{17}cm^{-2}}$,
confirming that the exact integration starting radius is not important for column densities
in the relevant range for significant attenuation of LyC photons.
In both cases we also see that 
there is a rapid upturn of the column density starting around ${\rm \sim 10^{18}cm^{-2}}$,
indicating the radial location of the beginning of self-shield and transition from a highly
ionized to an increasingly neutral medium.
The rapid ascent suddely flattens out at ${\rm \sim 10^{20}cm^{-2}}$,
sigalling the arrival of a largely neutral medium, coincidental with column density
similar to that of the damped Lyman alpha systems (DLAs). 
It is instructive to note that the transition from 
ionized to an increasingly neutral medium is halo virial radius (or halo mass) 
dependent, with a larger halo transitioning at a larger radius in units of its virial radius.
This indicates that the density of the ionizing front propagating into halos 
is halo mass dependent, suggesting that the common practice 
of using a constant density as a proxy for the density of ionization front
\citep[e.g.,][]{2000MiraldaEscude} could potentially be slightly extended,
although a more detailed analysis should be performed to assess this.

To devise an appropriate method to compute the effective cross section ${\rm A_{LL}}$
for LyC photons for a given halo,
it is useful to gain a more clear understanding of the physical meaning 
of ${\rm \lambda_{mfp,halo}}$.
For a line of sight cross area of size ${\rm \Delta A}$, 
if it is completely opaque to ionizing photons,
then the effective area for intercepting LyC photons would be just equal to ${\rm \Delta A}$.
For a cross area of size ${\rm \Delta A}$
that is not completely opaque to LyC photons,
one may define the effective area for intercepting ionizing photons 
${\rm \Delta A_{LL}}$, which is 
\begin{equation}
{\rm \Delta A_{LL} = \Delta A [1-\exp{(-N_{HI}\bar\sigma_{ion})}]},
\label{eq:ALL}
\end{equation}
\noindent
where ${\rm N_{HI}}$ is the column density integrated along that line of sight 
(not the radially integrated column density shown in the 
top-left panel of Figure \ref{fig:NFWcol}),
which is computed using 
${\rm N_{HI}(r)}$ and ${\rm x_{HI}(r)}$
that we have numerically obtained solving Eq (\ref{eq:NHIr},\ref{eq:local}).

Upon integrating the projected area of a halo,
we obtain the cumulative cross section for ionizing photons of a halo as a function of projected radius ${\rm r_p}$
\begin{equation}
{\rm A_{LL}(<r_p) = \int_0^{r_p}2\pi r_p^\prime [1-\exp{(-N_{HI}(r_p^\prime)\bar\sigma_{ion})}] dr_p^\prime}.
\label{eq:ALL}
\end{equation}
\noindent
The top-right panel of 
Figure \ref{fig:NFWcol} shows 
${\rm A_{LL}(<r_p)}$ in units of the virial area (${\rm\pi r_v^2}$) 
as a function of halo-centric radius in units of the virial radius ${\rm r_v}$
for the two halos shown in the top-left panel of Figure \ref{fig:NFWcol}.
To re-iterate a point made earlier,
the total effective cross section 
is larger for larger halos in units of the virial area,
shown quantitatively in 
the bottom-left panel of Figure \ref{fig:NFWcol}.
In the calculations performed involving the NFW profile, 
one needs to specify the concentration parameter ${\rm c}$,
which has been computed by a number of groups
\citep[e.g.,][]{2001Bullock, 2002Wechsler, 2016Angel, 2007Ricotti}.
We adopt the results of \citet[][]{2004Dolag}:
${\rm c=9.6 (M_h/10^{14}\msun)^{-0.10} (1+z)^{-1}}$;
the results obtained do not sensitively depend on slightly different formulae of ${\rm c}$
in the literature.

We compute ${\rm \lambda_{mfp,halo}}$ by
\begin{equation}
{\rm \lambda_{mfp,halo}^{-1} =\int_{M_{cut}}^\infty n(M_h) M_h \ln 10\ A_{LL}(M_h) d\log M_h}, 
\label{eq:lambdahalo}
\end{equation}
\noindent
where ${\rm A_{LL}(M_h)}$ is the total cross section of LyC photons
for a halo of mass ${\rm M_h}$;
${\rm n(M_h)}$ is the halo mass function at the redshift in question.
The bottom-right panel of 
Figure \ref{fig:NFWcol} shows cross section function, ${\rm n(M_h) M_h \ln 10\ A_{LL}(M_h)}$
(solid blue curve), 
its cumulative function ${\rm A_{LL,tot}(>M_h)}$ (dotted blue curve),
along with mass function, ${\rm n(M_h) M \ln 10}$(dashed red curve), as a function of ${\rm M_h}$.
We see that the cross section function is significantly flatter than
the halo mass function, due to the fact that the cross section 
in units of virial area is higher with increasing halo mass,
i.e., ${\rm A_{LL}(M_h)/M_h^{2/3}}$ correlates positively with ${\rm M_h}$,
shown in the bottom-left panel of Figure \ref{fig:NFWcol}.
Nonetheless, ${\rm A_{LL}}$ scales still sub-linearly with ${\rm M_h}$, causing  
${\rm n(M_h) M \ln 10 A_{LL}(M_h)}$
to increase with decreasing halo mass ${\rm M_h}$.

The ${\rm \Gamma-\lambda_{mfp}}$ relation in the standard $\Lambda$CDM model
for four cases of \\ ${\rm M_{cut}=(1.6\times 10^8, 5.8\times 10^7, 2.7\times 10^7, 8.6\times 10^6)\msun}$,
corresponding to a halo virial temperature cutoff of ${\rm T_{v,cutoff}=(10^4,5\times 10^3, 3\times 10^3, 1.4\times 10^3})$K,
are shown also in Figure \ref{fig:mfp} as the blue curves.
First of all our results affirm a general self-consistency between radiation field and ionization structures around halos in the $\Lambda$CDM model,
since the theoretically predicted relation (the blue curves) can go through this already tightly constrained parameter space.
This is a strong and unique support for the $\Lambda$CDM model with respect to its matter density
power spectrum (both amplitude and shape) 
on small scales corresponding to halo masses approximately in the range of $10^7-10^{10}\msun$.
It is noted that this constraint on matter power spectrum 
is based entirely on the consideration of the halos as ``sinks" of ionizing photons.
We point out the fact that ${\rm \lambda_{mfp,halo}}$ depends sensitively 
on the lower mass cutoff ${\rm M_{cut}}$ in the integral in Eq \ref{eq:lambdahalo},
as shown in the bottom-right panel of Figure \ref{fig:NFWcol}.
We show that this dependence provides a new, sensitive probe of the small-scale power in the cosmological model,
when confronted with measurements of ${\rm \tau_e}$.
It is useful to note that in computing ${\rm \lambda_{mfp,halo}}$ we have neglected
possible constribution due to collisional ionization in halos with virial
temperature significantly above $10^4$K.
Thus, our computed ${\rm \lambda_{mfp,halo}}$ is somewhat overestimated and our subsequent conclusion
drawn on small-scale power conservative.

\ifmnras

\begin{figure}
\centering
\includegraphics[width=3.5in, keepaspectratio]{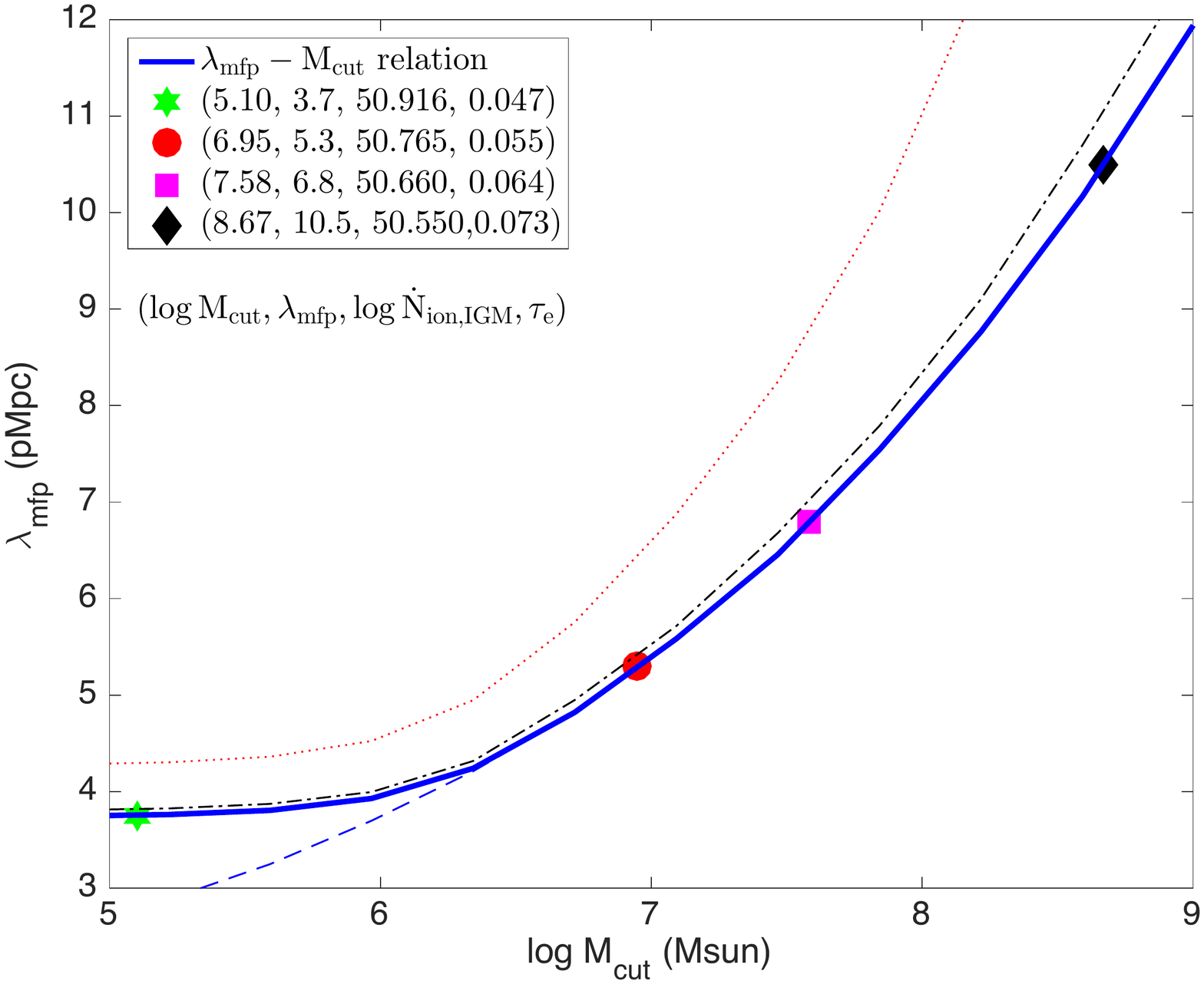}
\caption{
{\color{red}Left panel:}
shows ${\rm \lambda_{mfp}}$ as a function of the lower mass cutoff $M_{cut}$ 
in the integral in Eq \ref{eq:lambdahalo} (blue solid curve).
Also shown as symbols are four cases along the curve, 
with ${\rm (\log M_{cut}/\msun, \lambda_{mfp}/pMpc, \log\dot N_{ion,IGM}/cMpc^{-3}s^{-1}, \tau_e)}$
equal to 
$(5.10, 3.7, 50.916, 0.047)$ (green star),
$(6.95, 5.3, 50.765, 0.055)$ (red dot),
$(7.58, 6.8, 50.660, 0.064)$ (magenta square)
and  
$(8.67, 10.5, 50.550,0.073)$ (black diamond).
The thin blue dashed line is obtained when no entropy floor
due to that of mean cosmic gas is imposed.
The thin, black dot-dashed and red dotted curves
are obtained assuming there is no contribution from halos
with virial temperature greater than $3\times 10^5$K
and $3\times 10^4$K, respectively
}
\label{fig:mfpMcut}
\end{figure}

\else

\begin{figure}[!h]
\begin{center}
\includegraphics[width=5.5in, keepaspectratio]{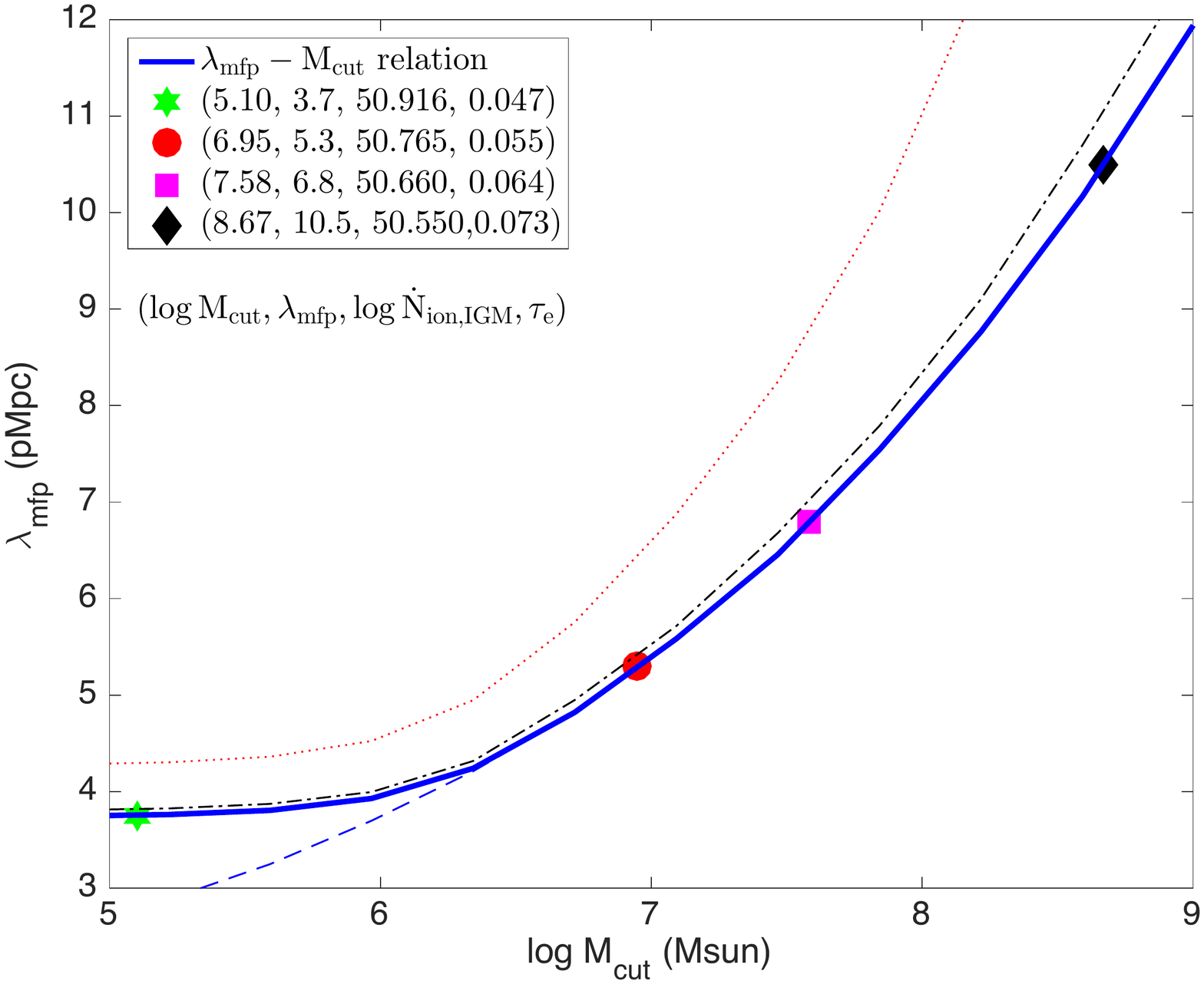}
\end{center}
\vskip -1.0cm
\caption{
shows ${\rm \lambda_{mfp}}$ as a function of the lower mass cutoff $M_{cut}$ 
in the integral in Eq \ref{eq:lambdahalo} (blue solid curve).
Also shown as symbols are four cases along the curve, 
with ${\rm (\log M_{cut}/\msun, \lambda_{mfp}/pMpc, \log\dot N_{ion,IGM}/cMpc^{-3}s^{-1}, \tau_e)}$
equal to 
$(5.10, 3.7, 50.916, 0.047)$ (green star),
$(6.95, 5.3, 50.765, 0.055)$ (red dot),
$(7.58, 6.8, 50.660, 0.064)$ (magenta square)
and  
$(8.67, 10.5, 50.550,0.073)$ (black diamond).
The thin blue dashed line is obtained when no entropy floor
due to that of mean cosmic gas is imposed.
The thin, black dot-dashed and red dotted curves
are obtained assuming there is no contribution from halos
with virial temperature greater than $3\times 10^5$K
and $3\times 10^4$K, respectively
}
\label{fig:mfpMcut}
\end{figure}

\fi

\ifmnras

\begin{figure}
\centering
\includegraphics[width=3.5in, keepaspectratio]{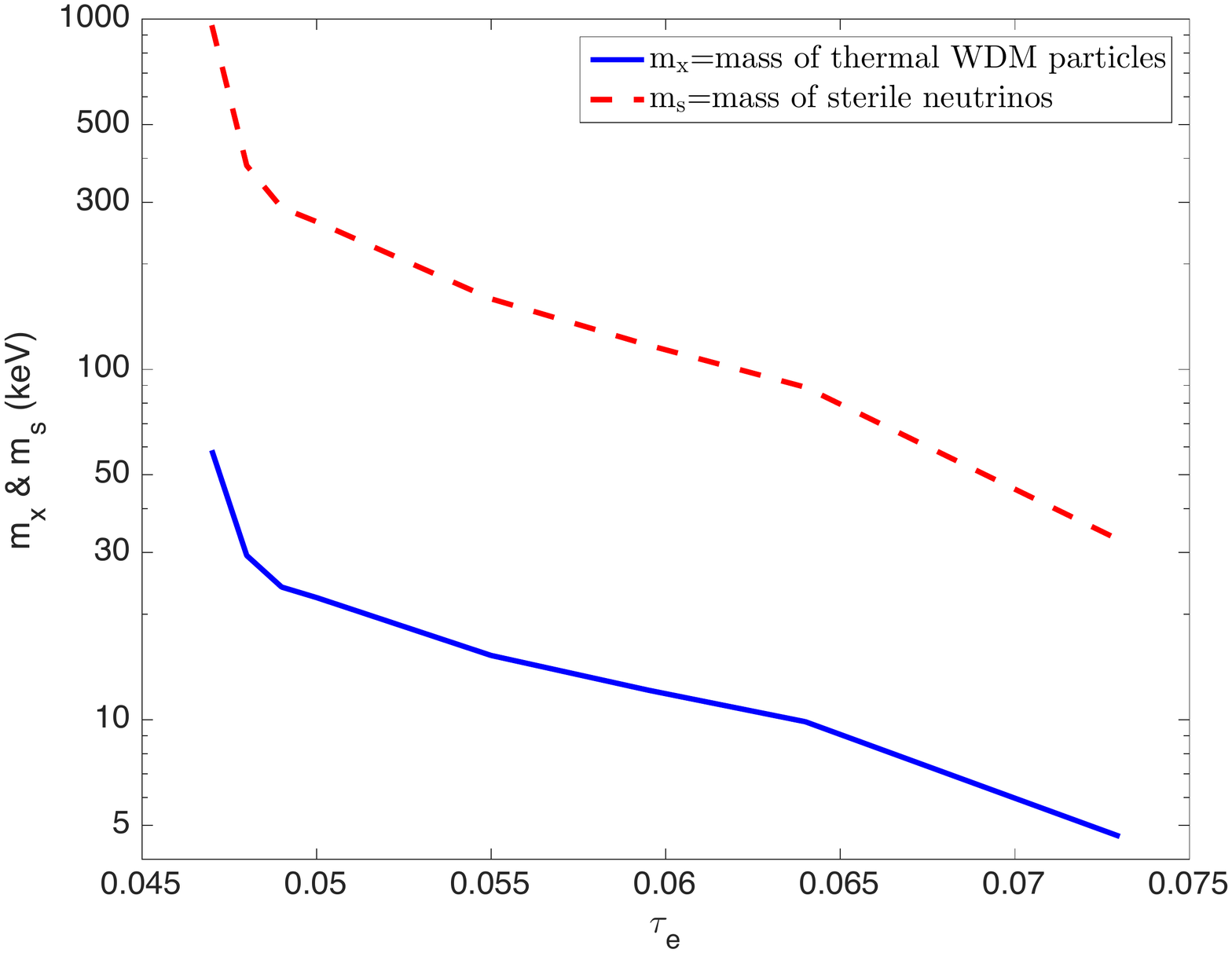}
\caption{
{\color{red}Left panel:}
shows the intrinsic 
}
\label{fig:mxtaue}
\end{figure}

\else

\begin{figure}[!h]
\begin{center}
\includegraphics[width=5.5in, keepaspectratio]{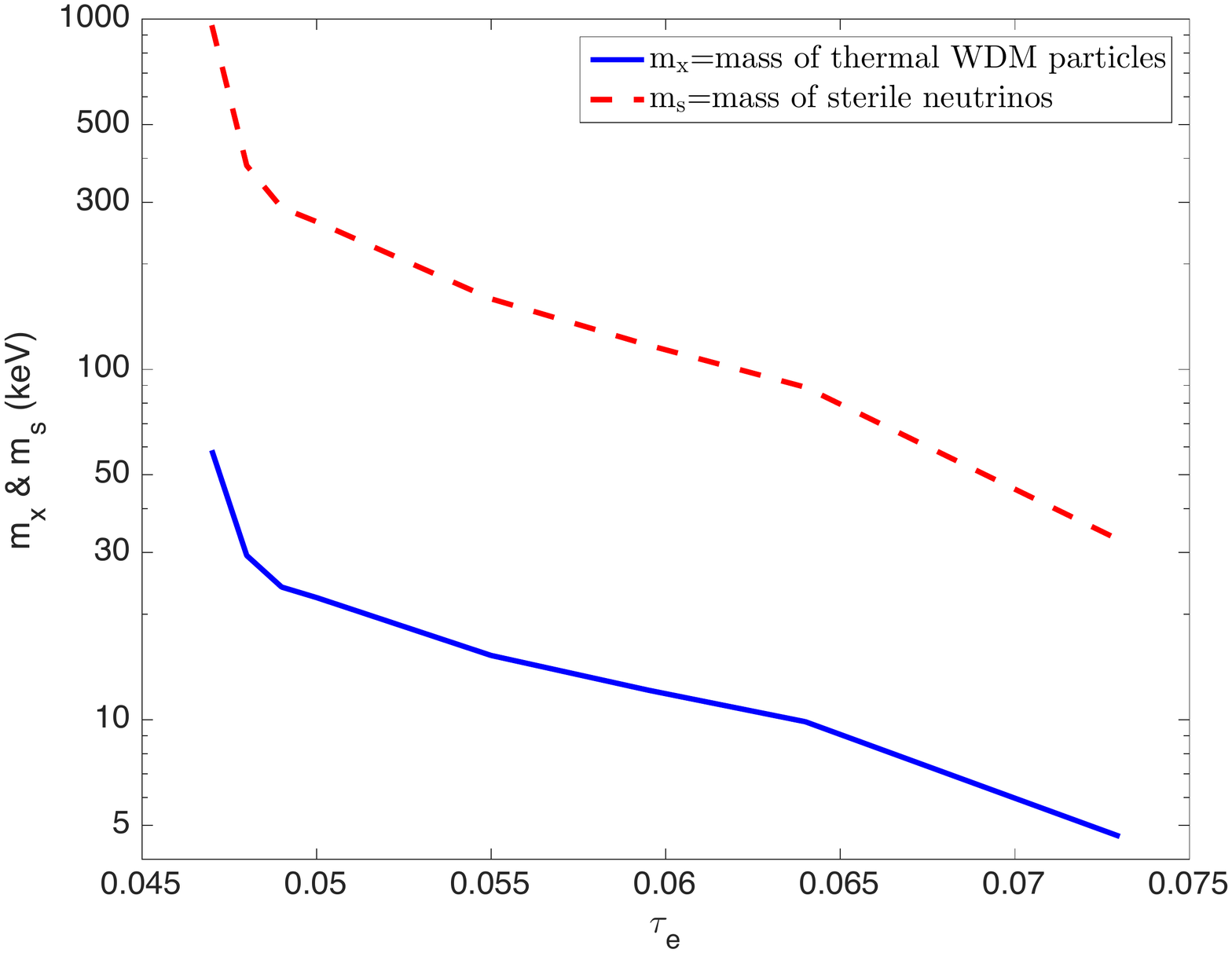}
\end{center}
\vskip -1.0cm
\caption{
shows the lower bound on the mass ${\rm m_x}$ 
of thermally produced warm dark matter particles 
as a function of ${\rm \tau_e}$ (blue solid curve).
Similarly, the red dashed curve shows 
the lower bound on the mass ${\rm m_s}$ 
of sterile neutrinos  
as a function of ${\rm \tau_e}$.
}
\label{fig:mxtaue}
\end{figure}

\fi

Figure \ref{fig:mfpMcut} shows
${\rm \lambda_{mfp}}$ as a function of the lower mass cutoff $M_{cut}$ 
in the integral in Eq \ref{eq:lambdahalo} (blue solid curve).
Shown as symbols are four cases along the curve, with \\
${\rm (\log M_{cut}/\msun, \lambda_{mfp}/pMpc, \log\dot N_{ion,IGM}/cMpc^{-3}s^{-1}, \tau_e)}$
equal to 
$(5.10, 3.7, 50.916, 0.047)$ (green star),
$(6.95, 5.3, 50.765, 0.055)$ (red dots),
$(7.58, 6.8, 50.660, 0.064)$ (magenta square)
and \\ 
$(8.67, 10.5, 50.550,0.073)$ (black diamond).
Each set of four numbers has the following relational meaning:
for a given measurement of ${\rm \tau_e}$,
the minimum required ionizing photon emissivity entering the IGM is 
${\rm \log\dot N_{ion,IGM}}$ in order for that ${\rm \tau_e}$ to be a possible solution, 
which in turn corresponds to a mean free path of ${\rm \lambda_{mfp}}$,
which can be achieved if the lower mass cutoff
of the halo mass function is 
${\rm M_{cut}}$. 
We see that the dependence of 
${\rm \lambda_{mfp}}$ on ${\rm M_{cut}}$ is significant,
which provides a new constraint on the small-scale power in the cosmological
model at a level that has hitherto been out of reach. 

The dependence of 
${\rm \lambda_{mfp}}$ on ${\rm M_{cut}}$ shown in Figure \ref{fig:mfpMcut} 
can be translated into a constraint on dark matter particles.
Here, we take warm dark matter as an example.
In the warm dark matter model the smoothing scale, 
defined as the comoving half-wavelength of the mode for which
the linear perturbation amplitude is suppressed by 2, is \\
\begin{equation}
{\rm R_s=0.48 ({\Omega_M/0.25})^{0.11}({h/0.7})^{-1.22}({m_x/{\rm keV}})^{-1.11} h^{-1}Mpc}
\label{eq:rs}
\end{equation}
\noindent
for a warm dark matter particle mass of ${\rm m_x}$ \citep[e.g.,][]{2005Viel},
which we adopt as a proxy for a sharp cutoff (or free-streaming scale of particles).
The equivalent free-streaming halo mass is then
\begin{equation}
{\rm M_s=5.8\times 10^{10} ({\Omega_M/0.3})^{1.33}({h/0.7})^{-4.66}({m_x/{\rm keV}})^{-3.33}\msun}.
\label{eq:Ms}
\end{equation}
\noindent
Given the dependence chain of 
${\rm \log M_{cut}}$ on ${\rm \lambda_{mfp}}$ on ${\rm \dot N_{ion,IGM}}$ on ${\rm \tau_e}$,
we obtain the lower bound on the mass ${\rm m_x}$ 
of thermally produced warm dark matter particles as a function of ${\rm \tau_e}$ 
shown as the blue solid curve in Figure \ref{fig:mxtaue}. 
The lower bound on the mass ${\rm m_x}$ of thermally produced warm dark matter particles
can be translated similarly to a lower bound constraint on the mass ${\rm m_s}$ of 
sterile neutrinos produced via active-sterile neutrino oscillations obeying 
approximately a generalized Fermi-Dirac distribution.
In this case, the effect of sterile neutrino is approximately
the same as for thermally produced warm dark matter by using the following expression to relate the two masses
\citep{1996Colombi, 2005Viel}: 
\begin{equation}
{\rm m_s = 4.46 keV ({m_{x} \over 1 keV})^{4/3} ({0.12 \over \Omega_M h^2})^{1/3}}.
\label{eq:k12}
\end{equation}
\noindent
The result is shown as the red dashed curve in Figure \ref{fig:mxtaue}.

The current best constraint on ${\rm m_x}$ based on Ly$\alpha$ forest 
is ${\rm m_x\ge 3.3keV (2\sigma)}$ \citep[][]{2013Viel},
improving upon earlier studies that generally
constrain ${\rm m_x\ge 0.5 - 1keV}$ \citep[e.g.,][]{2000Narayanan, 2001bBarkana, 2005Viel, 2006Abazajian}.
Combining with the $1\sigma$ upper limit used for $\Gamma$ in our calculations,
we find 
\begin{equation}
{\rm m_x \ge (15.1, 9.8, 4.6)keV\ at\ (1, 1.4, 2.2\sigma)\ C.L.},
\label{eq:k12}
\end{equation}
\noindent
based on ${\rm \tau_e=0.055\pm 0.009}$ and $+1\sigma$ on $\Gamma$.
The corresponding constraint on sterile neutrino mass is 
\begin{equation}
{\rm m_s \ge (161, 90, 33)keV\ at\ (1, 1.4, 2.2\sigma)\ C.L.},
\label{eq:k12}
\end{equation}
\noindent
which basically rules out, for example,
$7$keV sterile neutrino dark matter model 
\citep{2014Bezrukov, 2014Park, 2014Abazajian}. 
The lower bound placed on warm dark matter particle mass (or in general, on the small-scale power)
hinges on the assumption that dark matter halos make up
the bulk of the Lyman limit systems at $z=5.7$.
Are there possible caveats with respect to this assumption?  Let us examine this.

Under a physically plausible scenario of stellar reionization,
there are possibly two additional kinds of (significantly) neutral 
systems to serve as Lyman limit systems to contribute to the absorption of LyC photons.
The first kind is neutral regions that envelope the expanding HII regions.
Let us suppose that each HII region that is expanding has 
a radius of ${\rm R}$ and the neutral region surrounding it has a thickness
of ${\rm\Delta R}$.
Analysis of the Ly$\alpha$ forest at $z=5.7$ indicates 
a volume-weighted neutral fraction of the IGM 
${\rm f_{HI,V}\sim 0.9\times 10^{-4}}$ at $z=5.7$ \citep[][]{2006Fan}.
This provides a constraint on the possible size of ${\rm\Delta R}$:
\begin{equation}
{\rm\Delta R \le {f_{HI,V} R\over 3}}. 
\label{eq:DR}
\end{equation}
\noindent
The ionization front propagation speed at $z=5.7$ is 
\begin{equation}
{\rm v_{IF} = {F\over \bar n_H} = {\Gamma\over \bar\sigma\bar n} = 1.7\times 10^4 ({\Gamma_{-12}\over 0.31})({\bar\sigma\over 3.16\times 10^{-18}cm^2})^{-1}\kms},
\label{eq:vIF}
\end{equation}
\noindent
where ${\rm \bar n_H}$ is mean hydrogen number density at $z=5.7$.
Thus, the time it takes to sweep through the radial shell of thickness $\Delta R$ would be
\begin{equation}
{\rm \Delta t = {\Delta R\over v_{IF}} \le {f_{HI,V} R \over 3v_{IF}} = 9.4\times 10^3 ({R\over 5.3pMpc}) ({f_{HI,V}\over 0.9\times 10^{-4}}) ({\Gamma_{-12}\over 0.31})^{-1}({\bar\sigma\over 3.16\times 10^{-18}})~yrs}.
\label{eq:Dt}
\end{equation}
\noindent
Thus, for any reasonable values of the parameters involved,
${\rm\Delta t}$ is much shorter than the Hubble time at $z=5.7$ (which is about 1~Gyr).
This suggests that such a configuration is highly unlikely.
Note that our assumption that these shells surround spherical HII regions
is not necessary but only for the ease of illustration.
If these spherical shells are replaced by pancaky bridges or filamentary bridges
between (or connecting) HII regions,
the results and conclusions based on the above analysis remain largely the same,
as long as the size of these pancakes or filaments are on the same order
of $\sim 10$pMpc;
in terms of our conclusion reached, even for a size of $1000$pMpc, our conclusion
remains unchanged.

The second kind of possible neutral regions may be comprised of patches
of neutral islands in the voids that are last reionized.
We approximate them as opaque spheres with a radius of
${\rm r_{void}}$ and a mean separation between them of ${\rm d_{void}}$, 
which can be related to the observed ${\rm f_{HI,V}}$:
\begin{equation}
{\rm {4\pi\over 3} r_{void}^3 d_{void}^{-3} \le f_{HI,V}}.
\label{eq:void}
\end{equation}
\noindent
The mean free path to LyC photons due to these islands would be
\begin{equation}
{\rm \lambda_{mfp,void} = {d_{void}^3 \over \pi r_{void}^2} \ge ({4\over 3})^{2/3}\pi^{-1/3} d_{void} f_{HI,V}^{-2/3} = 412 d_{void} ({f_{HI,V}\over 0.9\times 10^{-4}})^{-2/3}}. 
\label{eq:void}
\end{equation}
\noindent
The typical separations of voids, i.e., ${\rm d_{void}}$,
has to be on the order of the clustering scale of galaxies, which 
is about $4-5$cMpc \citep[e.g.,][]{2010Ouchi},
or larger.
This suggests that ${\rm \lambda_{mfp,void} \ge 245}~$pMpc at $z=5.7$,
implying that possible, to-be-last-reionized neutral islands in voids
do not contribute much to the mean free path of LyC photons at $z=5.7$.

We thus conclude that halos likely contribute predominantly to the mean free
path of LyC photons at $z=5.7$ (likely at all lower redshifts as well, for that matter).
Finally, we note that for simplicity we have 
adopted the assumption of sphericity of gas distribution in and around 
halos in question.
Any deviation from sphericity would result in a reduction in cross section
hence a more stringent demand for more small scale power.
In addition, we note that baryonic fraction may be lower than the mean universal fraction.
Furthermore, some gas in large halos with virial 
temperature higher than $\sim 10^4$K may be heated up to remove itself from the HI category.
To give a sense of the magnitude of this effect
we show in Figure \ref{fig:mfpMcut} 
two additional cases where 
we assume that halos with virial temperature greater than $3\times 10^5$K
(thin, black dot-dashed curve) 
and $3\times 10^4$K (thin red dotted curve), respectively,
do not contribute to ${\rm \lambda_{mfp}}$. 
We see a significant effect; numerically, 
to attain ${\rm \lambda_{mfp}=(5.3, 6.8, 10.5)pMpc}$ in order to yield
${\rm\tau_e=(0.047, 0.055, 0.064, 0.073)}$, respectively,
the required ${\rm \log M_{cut}}$ changes
from $(8.67, 7.58, 6.95)$ for no upper cutoff 
to  $(8.54, 7.51, 6.89)$ for upper cutoff of virial temperature of $3\times 10^5$K,
to $(7.92, 7.07, 6.51)$ for upper cutoff of virial temperature of $3\times 10^4$K.
Moreover, internal ionizing radiation may reduce the HI fraction.
Therefore, our assumptions and derived limits on small-scale power
and on dark matter particle mass are all on the conservative side.

\section{Discussion}

\subsection{Rapid Reionization Towards $z=5.7$}

The intrinsic emissivities of LyC photons
at $z=5.7$ and $z=6$ are almost identical.
We can use this fact to outline the nature of percolation of HII regions 
near the end of the reionization.
We first note that we find that the 
theoretically derived relation of ${\rm \Gamma-\lambda_{mfp}}$ at $z=6$ is nearly identical 
to that at $z=5.7$ at the visual resolution of eye when overplotted in Figure \ref{fig:mfp}.
It means, if the universe were in the post-overlap regime already at $z=6$,  
its volume-weighted neutral fraction ought to be similar to that at $z=5.7$.
In other words, ${\rm \lambda_{mfp}}$ due to halos (mostly) based on $\Lambda$CDM model and emissivity at $z=6$ 
can easily accommodate a transparent universe similar to the one observed at $z=5.7$.
The observations indicate otherwise: 
${\rm f_{HI,V}\sim 0.9\times 10^{-4}}$ at $z=5.7$ versus
${\rm f_{HI,V}> 2\times 10^{-4}}$ at $z=6$ \citep[][]{2006Fan}.
Thus, the universe is not fully ionized at $z=6$ in the way of 
imposing a smaller 
${\rm \lambda_{mfp}}$ hence a lower ${\rm \Gamma}$ for a given ${\rm \dot N_{ion,IGM}}$. 
The likely, perhaps only, consistent solution would be that
HII regions have not overlapped at $z=6$ so that neutral patches in the IGM (not in the halos)
render ${\rm \lambda_{mfp}}$ much lower than  
the notional ${\rm \lambda_{mfp,IGM}}$ 
and 
${\rm \lambda_{mfp,halo}}$ in the post-overlap epoch.
The inferred value of $\Gamma_{-12}<0.02$ at $z=6$ (based on Ly$\gamma$ absorption)
\citep[][]{2002Cen, 2006Fan}
suggests that ${\rm \lambda_{mfp}}$ at $z=6$ is an order of magnitude
lower than that at $z=5.7$.
This is clear and fairly direct evidence that the percolation of HII regions is not yet complete at $z=6$,
indicating that the universe is in a rapid transitory phase from $z=6$ to $z=5.7$
clearing up some of the last neutral patches that dominate the mean free path,
in a monotonic and irriversible process.
Topologically, this indicates that HII regions transition from a set of isolated islands
at $z=6$ to a connected network of swiss-cheese-like HII region at $z=5.7$.

This expected rapid reionization process is consistent with 
and required by the necessary small values of ${\rm \lambda_{mfp}\le 6.8pMpc}$ at $z=5.7$
to achieve ${\rm \tau_e\le 0.064}$, 
which in turn requires contribution from minihalos (those with virial temperature 
less than $10^4$K or virial mass less than $1.6\times 10^8\msun$ at $z=5.7$).
Gas in minihalo, when exposed to ionizing photons,
responds dynamically by slowly evaporating through the action of thermal pressure 
of photoheated gas.
\citet[][]{2005Iliev} show that it takes about $100-200$Myr to photoevaporate 
a minihalo of mass $10^7\msun$ at $z=9$.
This process is expected to take longer for more massive minihalos.
In our case,  a minihalo of mass $10^7\msun$ is relevant for ${\tau_e=0.055}$ 
(see the red dot in Figure \ref{fig:mfpMcut});
for ${\tau_e=0.064}$ minihalos of mass $1.6\times 10^8\msun$ would be relevant 
(see the magenta square in Figure \ref{fig:mfpMcut}).
Thus, it is probably true that, for the range of interest,
the time scale taken for photoevaporation of relevant minihalos
is $100-200$Myr or longer.
We note that the universal age difference from $z=6$ to $z=5.7$ is $63$Myr,
from $z=7$ to $z=5.7$ is $231$Myr.
We see in Figure \ref{fig:green} that the neutral fraction at $z=7$ is about $40\%$,
meaning about 40\% of minihalos have not yet been exposed to ionizing radiation at $z=7$.
Thus, it is probable that a significant fraction, perhaps a large majority,
of minihalos have not lost gas in their inner regions (that actually contribute
to the mean free path of LyC photons) by $z=5.7$,
permitting the possibility 
that they contribute significantly to the mean free
path of LyC photons, if necessary.

\subsection{On ${\rm f_{esc}}$ of Galaxies at Epoch of Reionization}

Using Eq \ref{eq:fesc}, the four points (represented by the four symbols)
in Figure \ref{fig:mfpMcut} give \\
${\rm f_{esc}=(20.7, 14.6, 11.5, 8.9)\%}$,
in order to arrive
at the reionization solutions constrained by the state of the IGM at $z=5.7$ with 
${\rm \tau_{e}=(0.047, 0.055, 0.064, 0.073)}$, respectively.

This required ${\rm f_{esc}}$ based on the observed state of the IGM at $z=5.7$
is consistent with computed ${\rm f_{esc,comp}=10-14\%}$
based on state-of-the-art high resolution cosmological radiation
hydrodynamic simulations of dwarf galaxies at the epoch of reionization
of \citet[][]{2014Kimm}.
We point out that the upper value (14\%) includes contributions from runaway OB stars.
It is noteworthy that ${\rm f_{esc,comp}}$ is effectively a measure of the porosity 
of the interstellar medium, where LyC photons escape
through transparent holes into the IGM.
Therefore, a correct treatment/implementation of supernova feedback
is essential, as is in \citet[][]{2014Kimm} 
but not in any other simulations that the author is aware of.
Including Wolf-Rayet stars for Pop II stellar population, 
which empirically are much more abundant in local metallicity
environment that is expected for galaxies at the epoch of reionization,
may further increase the ratio of LyC photons to FUV photons, 
i.e., ${\rm \xi_{ion}}$, thus lessen the requirement for a high ${\rm f_{esc}}$.
Thus, it seems that the stellar emissivity observed is adequate for 
maintaining the state of the IGM in terms of global and local ionization balance.
It should be noted that these changes have no effect on solutions of reionization history
that we have obtained, which depends directly on ${\rm \dot N_{ion,IGM}}$.

\subsection{Dichotomy in the Evolution of Lyman Alpha Emitters $z>6$}

In Figure \ref{fig:zoom} we see
that solutions without Pop III contributions
require $\chi=(0.7,2.2,3.6)$ for 
${\rm \tau_{e}=(0.055, 0.064, 0.073)}$, respectively.
In general, the solutions even with Pop III contributions
requires $\chi>0$ as long as ${\rm \tau_{e}\ge 0.052}$.
We note that the overall ${\rm f_{esc}}$ tends to 
correlate with the porosity of the ISM, while individual 
${\rm f_{esc}}$ is strongly dependent on the line of sight of 
the observer \citep[e.g.,][]{2015bCen}.
A positive $\chi>0$ is physically consistent with 
the expectation that smaller galaxies, having shallower gravitational potential wells,
may be more susceptible to feedback processes from supernovae and have more porous ISM.
Simulation results are consistent with this expected trend \citep[e.g.,][]{2014Kimm}.

Is there observational evidence that 
the escape of Ly$\alpha$ and of LyC photons are both correlated with ISM posority?
\citet[][]{2013Jones} find 
an interesting trend of lower covering fractions of low-ionization gas for galaxies with strong 
Ly$\alpha$ emission, 
providing evidence for a reduction in the average HI covering fraction
(hence an increase in the escape fraction of ionizing radiation) is correlated with 
increase in Ly$\alpha$ emission.
\citet[][]{2003Shapley}
find that the blueshifts of interstellar absorption lines in LAEs and LBGs
are similar at $\sim -200\kms$, suggesting that the velocity of outflows
in LAEs and LBGs are comparable.
But their study also reveals
a trend that Ly$\alpha$ EW increases with decreasing
${\rm \Delta v_{em-abs}}$ in the EW range of $-15$ to $+50$\AA. 
Furthermore, they confirm that 
${\rm \Delta v_{Ly\alpha}}$ of LAEs is systematically smaller than the values of LBGs,
with $\Delta v_{Ly\alpha}$ of about $200\kms$ for LAEs compared to about $400\kms$ for LBGs.
Moreover, they clarify that ${\rm \Delta v_{Ly\alpha}}$ decreases with increasing EW of Ly$\alpha$.
Recently, \citet[][]{2014Shibuya}
find an anti-correlation between Ly$\alpha$ EW and the covering fraction
estimated from the depth of absorption lines, 
which is an indicator of average neutral hydrogen column density. 
Their results support the idea that neutral column density is a key quantity determining Ly$\alpha$ emissivity,
consistent with the notion that the escape of LyC and Ly$\alpha$ is correlated with each other
and due to lower column density holes in the ISM.
The combination of these facts leads one to conclude that
the Ly$\alpha$ velocity offset is positively correlated 
with ${\rm N_{HI}}$ and negatively correlated with EW,
exactly predicted from results based on Ly$\alpha$ radiative transfer calculations
\citep[e.g.,][]{2010Zheng}.
None of these properties concerning Ly$\alpha$ emission can
be attributed to differences in the outflow velocity, which do not appear to exist between LAEs and LBGs.
Taken together, intrinsically, one would have expected then that the
escape of Ly$\alpha$ photons should be made easier with increasing redshift; 
i.e., both the ratio of Lyman alpha emitters to overall galaxy population 
at a chosen Ly$\alpha$ EW or 
the overall Ly$\alpha$ luminosity to FUV luminosity ratio as a whole
are expected to increase with redshift beyond $z=5.7$.

Such an expectation is not borne out with observations.
At some EW cuts, observations have consistently found 
that the fraction of LAEs out of LBGs 
decreases by a significant factor from redshift $z=6$ to $z=8$
\citep[e.g.,][]{2013Treu,2014Vanzella,2014Faisst,2014Schenker,2014Tilvi,2016Furusawa}.
This observational evidence strongly suggests that the intergalactic medium 
may have increasingly diminished the observability of the Ly$\alpha$ from $z\sim 6$ to $z\sim 8$,
consistent with the rapid reionization picture depicted in Figure \ref{fig:green}).
Physically, this is due to the fact that significantly neutral IGM 
limits the size of Stromgren sphere around galaxies \citep[][]{2000Cen}. 
\citet[][]{2014Caruana} 
conclude that the neutral fraction of the IGM at $z \sim 7$ to be $\sim 0.5$,
which would be consistent with our computed model shown in 
Figure \ref{fig:green}).

On the other hand, even if the IGM is indeed masking the appearance of the Ly$\alpha$ emission
for most, relatively low luminosity galaxies at the epoch of reionization,
for rare, very luminous galaxies (which each are also likely clustered with other 
galaxies) with large Stromgren spheres,
their Ly$\alpha$ emission lines may be unaffected or possibly enhanced (given $\chi>0$),
under suitable conditions.
A corroborative or confirmative piece of evidence for this may be that,
if a strong Ly$\alpha$ line is detected,
the emission region could, but not necessarily required to,
be compact spatially and in velocity space due to lack of scattering.
There are observational indications that this may in fact be the case.
\citet[][]{2015Sobral}
observe a luminous Ly$\alpha$ source (CR7)
with luminosity of $10^{43.93\pm 0.05}$ erg/s at $z=6.6$ (the most luminous Ly$\alpha$ emitter ever found at $z>6$)
but with a narrow FWHM of $266\pm 15 \kms$. 
\citet[][]{2016Hu} detect
a luminous Ly$\alpha$ emitting galaxy, COLA1,
with luminosity of $10^{43.9}$ erg/s at $z=6.593$.
COLA1 shows a multi-component Ly$\alpha$ 
profile with a blue wing, suggesting a large and highly 
Stromgren sphere perhaps well extending into the infall region.
\citet[][]{2015Matthee} have argued that there is little
evolution in the luminosity function of the most luminous LAEs at these redshifts, 
suggesting that these objects
lie in large HII regions and protect themselves from 
changes in IGM neutral fraction, 
consistent with the expectation, at least in principle.
More pinpointed analysis will be desirable in this respect,
combining reionization simulations with detailed radiative transfer of Ly$\alpha$ photons.

In summary, we expect that there is a dichotomy in the evolution of Ly$\alpha$ emitting galaxies.
For relatively low Ly$\alpha$ luminosity galaxies, their emission lines will be
progressively diminished with increasing redshift
due to the increasingly neutral IGM beyond $z\sim 6$.
On the other hand, for the most luminous Ly$\alpha$ emitters, under suitable conditions,
their Stromgren spheres are large enough to allow their Ly$\alpha$ line to escape unscathed
by the neutral IGM. 
Both are consistent with present tentative observational evidence.

\section{Conclusions}

We utilize the joint observations of the Ly$\alpha$ forest,
the mean free path of ionizing photons ${\rm \lambda_{mfp}}$, the luminosity function of galaxies
and the total electron scattering optical depth $\tau_e$, 
and theoretical insight on a relation between matter power spectrum and ${\rm \lambda_{mfp}}$,
to perform a detailed analysis 
of the solutions of cosmic reionization history that satisfy 
the observed boundary conditions of the IGM at $z=5.7$.
We summarize results and conclusions.

(1) A theoretical relation between the mean free path and ionization rate at $z=5.7$,
requiring only the matter power spectrum, is derived.
More scale power on $10^6-10^9\msun$ scales leads to lower mean free path. 

(2) A negative relation is found between the minimum effective ionizing photon emissivity for the IGM at $z=5.7$
and the electron scattering optical depth ${\rm \tau_e}$.
A higher emissivity is coupled with a less steep increase of ionizing photon
escape fraction with increasing redshift,
resulting in a later reionization episode hence a lower ${\rm\tau_e}$.

(3) The minimum required mean escape fraction of ionizing photons from galaxies at $z=5.7$ is found to be
${\rm f_{esc}=(20.7, 14.6, 11.5, 8.9) \left({\xi_{ion}\over 10^{25.2}}\right)^{-1}\%}$
for ${\rm \tau_{e}=(0.047, 0.055, 0.064, 0.073)}$, respectively,
where ${\rm \xi_{ion}}$ is the ratio of ionizing photo production rate (in ${\rm cMpc^{-3}~s^{-1}}$)
to FUV spectral density (in ${\rm erg~s^{-1}~Hz^{-1}~cMpc^{-3}}$).
The escape fraction is predicted to increase with increasing redshift,
with the rate of increase required higher for higher ${\rm\tau_e}$.

(4) While there is a family of possible solution,
the 50\% ionization fraction redshift lies
in a relatively narrow range of $z=6.5-7.5$ for ${\rm \tau_{e}=0.050-0.082}$. 
The late reionization suggests that relatively low luminosity Ly$\alpha$ emitters beyond $z=6$,
incapable of carving out a sufficiently large Stromgren sphere, 
will be increasingly diminished,
although the most luminous Ly$\alpha$ emitters 
may possess a large enough Stromgren sphere to allow 
unimpeded transmission of their Ly$\alpha$ lines,
possibly characterized by compact spatial or velocity extent.

(5) Topologically, reasonable arguments lead to the picture that 
the universe transitions from a set of isolated HII bubbles of typical individual sizes
probably no greater than $1$pMpc at $z=6$ to a 
set of isolated neutral islands centered on halos that
are embedded in one connected of HII region at $z=5.7$.

(6) A positive relation is found between ${\rm \tau_e}$
and the maximum mean free path of ionizing photons at $z=5.7$.
The outcome comes about because the product of 
the free path and emissivity of ionizing photons 
at $z=5.7$ is constrained by the observed Gunn-Peterson optical depth.
The maximum mean free path at $z=5.7$ is 
${\rm (3.7, 5.3, 6.8, 10.5)pMpc}$ in order to yield
${\rm\tau_e=(0.047, 0.055, 0.064, 0.073)}$, respectively.
We do not find it possible to find a reionization solution with ${\rm\tau_e<0.047}$
that satisfies all observed conditions. 

(7) The electron scattering optical depth ${\rm \tau_e}$ thus provides a constraint on 
the mean free path, which in turn yields a new and powerful
constraint on the matter power spectrum on $10^6-10^9\msun$ scales at $z=5.7$. 
With the latest Planck measurements of ${\rm \tau_e = 0.055 \pm 0.009}$,
we can place an upper limit of $(8.9\times 10^6, 3.8\times 10^7, 4.2\times 10^8)\msun$ 
on the cutoff mass of the halo mass function,
or equivalent a lower limit on warm dark matter particle mass ${\rm m_x \ge (15.1, 9.8, 4.6)keV}$
or on sterile neutrino mass ${\rm m_s \ge (161, 90, 33)keV}$ in the warm dark matter model, at $(1, 1.4, 2.2)\sigma$
confidence level.

(8) It is clear that a solution to the missing satellite problem \citep[][]{1999Klypin, 1999Moore} 
is unattainable via the route of warm dark matter particle origin,
because of the strong constraint on the upper bound on dwarf halo mass 
of $\le 4.2\times 10^8\msun$ at $2.2\sigma$ found. 


\medskip
I thank Xiaohui Fan, Jordi Miralda-Escude, Graca Rocha and Hy Trac for helpful discussion.
I also thank an anonymous referee for useful and constructive comments.
This work is supported in part by grants NNX12AF91G and AST15-15389.









\end{document}